\documentclass[11pt,a4paper]{article}
\usepackage{amsmath}
\usepackage{amsfonts}
\usepackage{amssymb}
\usepackage{graphicx}
\usepackage{bm,bbm}

\usepackage{indentfirst}
\usepackage[top=2cm,bottom=3cm,left=2cm,right=2cm]{geometry}

\DeclareMathOperator*{\Tr}{Tr}

\newcommand{\ket}[1]{|#1\rangle}
\newcommand{\braket}[2]{\langle #1|#2\rangle}
\newcommand{\ketbra}[2]{|#1\rangle\langle #2|}

\title{Random unitary matrices associated to a graph}

\author{Pawe{\l} Kondratiuk$^1$\footnote{Present address: Institute of Theoretical Physics, Faculty of Physics, University of Warsaw, Ho\.{z}a 69, 00-681 Warsaw, Poland}, Karol \.{Z}yczkowski$^{2,3}$}

\begin{document}

\maketitle

{\small
$^1${Faculty of Physics, 
Warsaw University of Technology, Koszykowa 75, PL-00-662 Warsaw, Poland}

$^2${Institute of Physics, Jagiellonian University, ul. Reymonta 4, 30-059 Krak{\'o}w, Poland}

$^3${Center for Theoretical Physics, PAS, Al. Lotnik{\'o}w 32/44, 02-668 Warszawa, Poland}
}

\bigskip


\begin{abstract}

We analyze composed quantum systems consisting of $k$ subsystems, 
each described by states in the $n$-dimensional Hilbert space. 
Interaction between subsystems can be represented by a graph, 
with vertices corresponding to individual subsystems and 
edges denoting a generic interaction, modeled by random unitary matrices of order $n^2$.
The global evolution operator is represented by a unitary matrix of size $N=n^k$.
We investigate statistical properties of such matrices
and show  that they display spectral properties characteristic to 
Haar random unitary matrices provided the corresponding graph is connected.
Thus basing on random unitary matrices of a small size $n^2$
one can construct a fair approximation of large random unitary matrices of size $n^{k}$. 
Graph--structured random unitary matrices investigated here
allow one to define the corresponding structured ensembles of random pure states.
\end{abstract}

\section{Introduction}

Random unitary matrices can be applied 
to describe quantum chaotic scattering or an evolution operator for a periodic, time--dependent system, the 
corresponding classical dynamics is chaotic in the entire phase space \cite{Ha06}.
If the system possess no time-reversal symmetry the 
corresponding operators display statistical properties 
typical to {\sl circular unitary ensemble} ({\sl CUE}) of matrices distributed
according the the Haar measure on the unitary group \cite{Mehta}.

A random matrix typical to CUE is hence related to a common physical situation, 
in which there exist a generic, possibly unspecified, interaction 
between any two levels of the entire system. In a more general setup
of a multi-partite system this assumption corresponds thus to a typical interaction
between any pair of subsystems.

On the other hand in a broad class of quantum models studied in condensed matter or atomic physics the interaction acts only locally between neighbouring
particles on a prescribed lattice. If the exact Hamiltonian describing
such an interaction is unknown, one can mimic it by a random unitary matrix
which couples only a few selected subsystems. In this way we arrive at a
model of random unitary matrices associated to a graph or a network,
which will be introduced and analyzed in this work.

The model described precisely in the next section, is related 
to the ensemble of {\sl structured}
 quantum pure states associated with a graph investigated in
\cite{CNZ10,CNZ13}. These assumptions differ significantly 
from the model analyzed in \cite{VC04},
in which edges of the graph represent maximally entangled states of two qubits,
while the vertices represent deterministic local unitary gates or local measurements. 
A similar idea of an edge representing a maximally entangled state of
two particles was also used in a deterministic construction 
of projected entangled pair states \cite{VW+06},
while more general models of quantum networks 
were analyzed in \cite{PC+08,Pe+10b}.

We extend here the model introduced in \cite{CNZ10}
of a random unitary interaction represented by each vertex of the graph, 
but make the model symmetric, by assuming that in the subsequent time step
a similar random interaction takes place along each bond of the graph.
Thus the physical role of bonds and edges of a graph
is in sense similar, in an analogy to 
the construction of line--graphs \cite{PTK03}.

The main aim of this work is to introduce 
ensembles of structured random unitary matrices associated to a graph
and to investigate their basic properties.
We report here a key observation concerning the spectral 
statistics of such structured unitary matrices.
On one hand ensembles of matrices
related to non-connected graphs display Poisson--like spectra.
On the other hand, a typical connected graph leads to an ensemble 
with several properties characteristic to CUE, 
even though the interaction takes place locally
between the subsystems connected by a bond 
or belonging to a single vertex of a graph.

The paper is organised as follows. In the next
section two alternative versions of the scheme associating a
random unitary matrix to a graph are described.
Statistical properties of spectra of random matrices
corresponding to exemplary graphs are analyzed in section 3.
The distribution of eigenvectores of graph unitary matrices
is analyzed in section 4, while statistical properties of their
entries are discussed in section 5. 
Concluding remarks are presented in section 6, while some details concerning the
numerical computations are provided in the Appendix.

\section{Interactions associated to a graph and corresponding unitary matrices}

We are going to discuss a general case of a composite quantum system consisting of 
an arbitrary number $k$ of subsystems isolated from the environment.
For simplicity we shall assume here
that each subsystem is described in an $n$ dimensional Hilbert space, ${\cal H}_n$.
Hence the total dimension of the Hilbert space reads $N=n^{k}$
and the composite system is described by a state $|\psi\rangle$ in
the composite Hilbert space ${\cal H}_N = {\cal H}_n \otimes \dots \otimes {\cal H}_n$.

A Hamiltonian evolution operator 
can be represented by a global unitary matrix $U$ of order $N$.
Assume first that the time evolution of the composite system
can be decomposed into two time steps,
so the time evolution is given by a product of two matrices
\begin{equation}
U = W V.
\label{eq:mat_mul}
\end{equation}
Here $W$ and $V$ denote unitary matrices, which describe
both phases of the time evolution, which occurs sequentially, 
one after another.

The main assumption of the model is that the physical interactions 
taking place between certain subsystems has 
a topology which can be described by a graph.
To make the presentation more complete we shall
define two different schemes of representing the 
interaction by a graph. Although
some interaction patterns can be described equally well
using any one of the two constructions proposed,
in some cases only one of these two schemes is applicable,
what provides a motivation to describe both of them.

\subsection{A bond of a graph represents two coupled subsystems}

In the first approach we will 
assume that the total number of subsystems is even $k=2m$
and the interaction can be represented by an undirected graph $\Gamma_1$ 
consisting of $m$ bonds and $v$ vertices.
In general the graph needs not to be connected and 
we may allow loops and multiple connections between vertices.


In the first time step of the evolution a generic
interaction takes place independently in each vertex of the graph.
Such an interaction is described by a random unitary operator
$V^{(j)}$, where $j=1,\dots, v$ labels the vertices of the graph.
For instance, if the first vertex couples the subsystems labeled by 
$2$ and $3$ we shall write $V^{(1)}=V_{2,3}$.
The interaction in all vertices is thus described by a tensor product
\begin{equation}
V = V^{(1)} \otimes V^{(2)} \otimes \cdots \otimes V^{(v)}.
\label{eq:vprod}
\end{equation}

Each bond of a graph represents two subsystems interacting
jointly in the second time step. For instance, the first bond,
connecting subsystems labeled by $1$ and $2$,
will denote a generic interaction between them 
represented by a random unitary matrix $W_{12}$ of order $n^2$.
Hence the second time step is described by a unitary matrix
of the product form 
\begin{equation}
W =  W_{1,2} \otimes W_{3,4} \otimes \cdots \otimes W_{k-1,k} \; ,
\label{eq:wprod}
\end{equation}
where $W_{2j-1,2j}$
describes a generic bi-partite interaction 
corresponding to $j$-th bond of the graph.
Hence the entire, two--step time evolution reads $U=WV$, according to 
(\ref{eq:mat_mul}), where both unitary terms $W$ and $V$ have a tensor
product structure.
Observe that the tensor product symbols $\otimes$
present in (\ref{eq:vprod}) are taken with respect to different partitions
of the total Hilbert space as these occurring in Eq. (\ref{eq:wprod}),
so in general the operator $U$ does not posses a tensor product structure.

\begin{figure}
\centering
\includegraphics[width=0.58\textwidth]{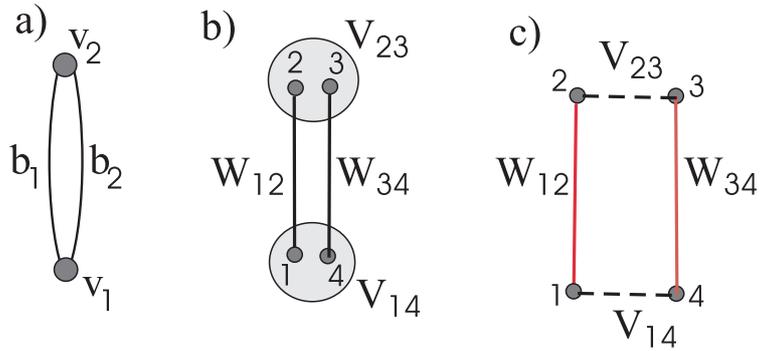}
\vskip -2.1cm
\caption{a) An exemplary graph with two vertices and two bonds describes a $4$--party system,
 b) construction of the corresponding random unitary matrix $U$ defined by Eq. (\ref{v4}); 
c) the same interaction described by Eq. (\ref{U22}) and 
represented by a two--color graph -- one color is represented by a solid (red) lines and
 the other by dashed (black) lines.}
\label{fig:gra13}
\end{figure}

To watch this construction in action consider a simple graph
consisting of two vertices, $v=2$ and two bonds between them, $m=2$.
This graph describes thus $k=2m=4$ subsystems, which are labeled here by 
$1,2,3,4$ -- see Fig. \ref{fig:gra13}.
The evolution operator constructed according to the rules
(\ref{eq:mat_mul},\ref{eq:vprod},\ref{eq:wprod}) reads thus
\begin{equation}
 U  \ =  \  \bigl( W_{1,2}  \otimes W_{3,4} \bigr) \;
\bigl(  V_{2,3}  {\tilde \otimes} V_{1,4} \bigr) ,
\label{v4}
\end{equation}
where the interaction along the bonds is given by random unitary matrices 
$W_{1,2}$ and $W_{3,4}$ of size $n^2$,
while interaction at the vertices is described by unitary
matrices $V_{2,3}$ and $V_{1,4}$ of the same size.
It is convenient to label unitary matrices $V$, describing 
interaction at a given vertex by its number, written in a superscript in brackets
or by the numbers of particle it includes placed in a subscript,
and freely switch between both conventions.
In the example described above 
one has $V^{(1)}=V_{2,3}$ and $V^{(2)}=V_{1,4}$.

The sign tilde over the second tensor product in Eq. (\ref{v4})
is put to emphasize that both tensor products are taken with 
respect to different partitions, so 
$U$ cannot be written as a tensor product of two local unitary matrices. 

\subsection{A vertex of a two-color graph represents a subsystem}
\label{sub2b}

In the second approach each of $v$ vertices of the graph describes a single subsystem,
hence the number of parties involved $k=v$ is arbitrary.
On the other hand, in this scheme we need to make some restrictions concerning the topology of the graph representing the interaction.
To describe physical interaction occurring in two moments of time
we will use two kinds of bonds, denoted in the graph by two different colours.
This construction is unambiguous provided the two--colour graph $\Gamma_2$
considered here
satisfies the following property: There exists a single bond of each colour 
linked to a given vertex (see Fig.2b-2e), or if there are more of them, 
(for example -- two red, solid bonds entering vertex $1$ in Fig.2f),
they are a part of a maximally connected (sub)graph of this colour (a triangle in this figure).

To present a formal definition of this property
we will use notation of the graph theory.
A {\sl clique} in the graph
is defined as a subset of vertices connected to each other.
Let $Q = \{q_i\}_{i=1}^k$ denotes the set of vertices of the graph.
A partition of the set $Q$, given by 
any set of its mutually exclusive and collectively exhaustive subsets 
is denoted by $\Pi(Q)$.
Our requirement concerning the graph $\Gamma_2$ is then equivalent to 
an assumption that its vertices can be divided into two 
partitions, $\Pi_1(Q)$ and $\Pi_2(Q)$, both of which 
consist of separate cliques only.
Each partition is represented
on the graph by bonds of a certain color.
Hence we introduce two sets of bonds, $B_1$ and $B_2$,
and represent the evolution operator by two graphs of interactions, 
$G_1 = \{Q, B_1\}$ and $G_2 = \{Q, B_2\}$

The unitary operator $U$ of the entire system, describing the two--step 
time evolution, can be therefore expressed as 
\begin{equation}
U = \left( \bigotimes^{\  }_{\pi \in \Pi_2(Q)} W_\pi \right) 
   \left({\tilde \bigotimes}_{\pi \in \Pi_1(Q)} V_\pi \right),
\label{U22}
\end{equation}
where operator $V_\pi$ (or $W_\pi$) acts on the particles from the subset $\pi$,
and 
the tilde over the sign $\otimes$ in the second term is
used to emphasize that the tensor products are taken with 
respect to different partitions.
The size of a component unitary matrix $V_\pi$ is a function of the number $\#\pi$ of the particles in the subset $\pi$  as $\dim V_\pi =  n^{\#\pi}$.
Note that the operator (\ref{U22})
is now uniquely determined by a two--color graph 
$\Gamma_2 = \{Q, \{B_1, B_2\}\}$ (consisting of a set of vertices and two sets of edges).

\begin{figure}
\centering
\includegraphics[width=0.16\textwidth]{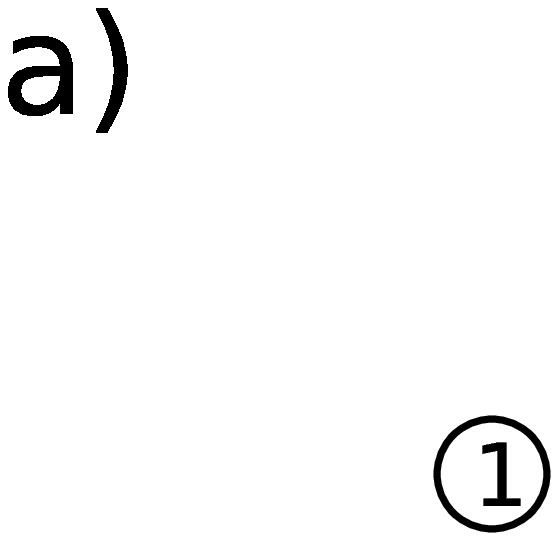}
\includegraphics[width=0.16\textwidth]{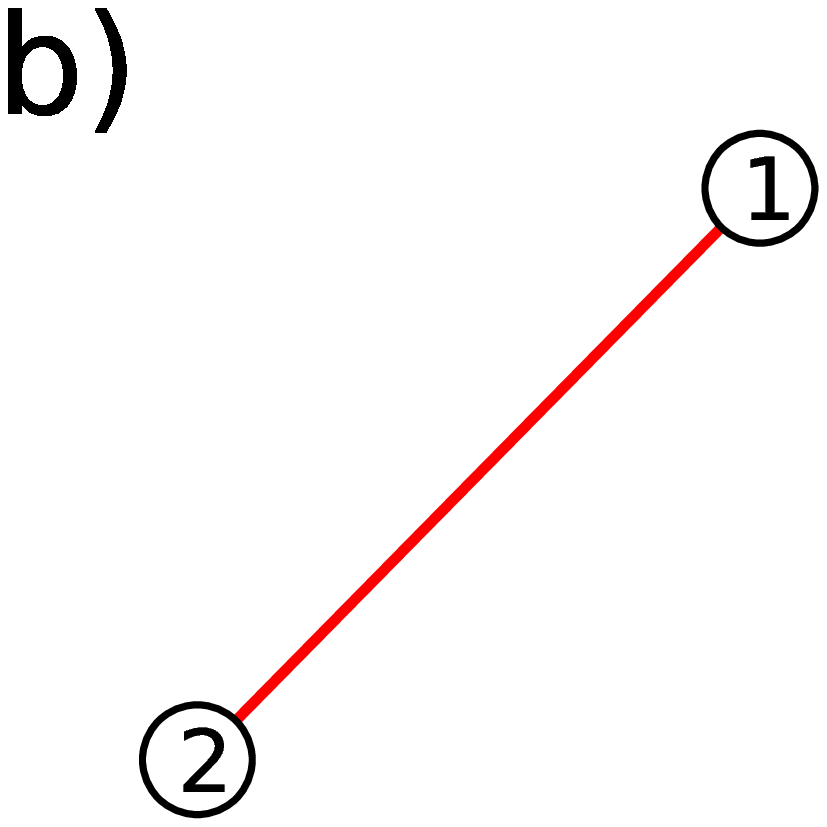}
\includegraphics[width=0.16\textwidth]{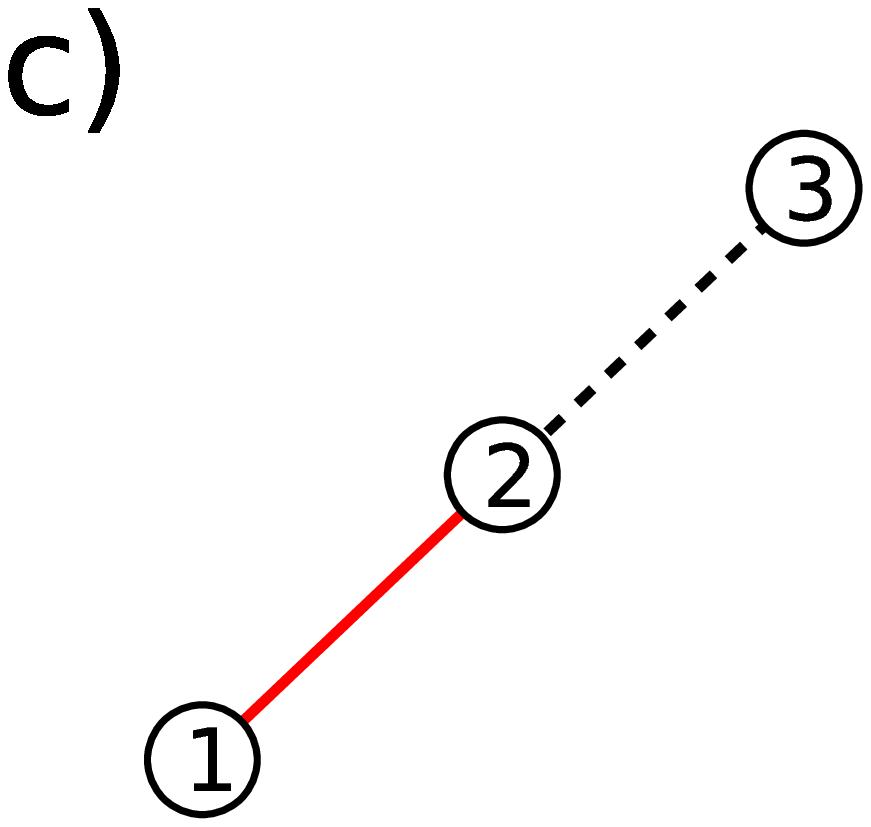}
\includegraphics[width=0.16\textwidth]{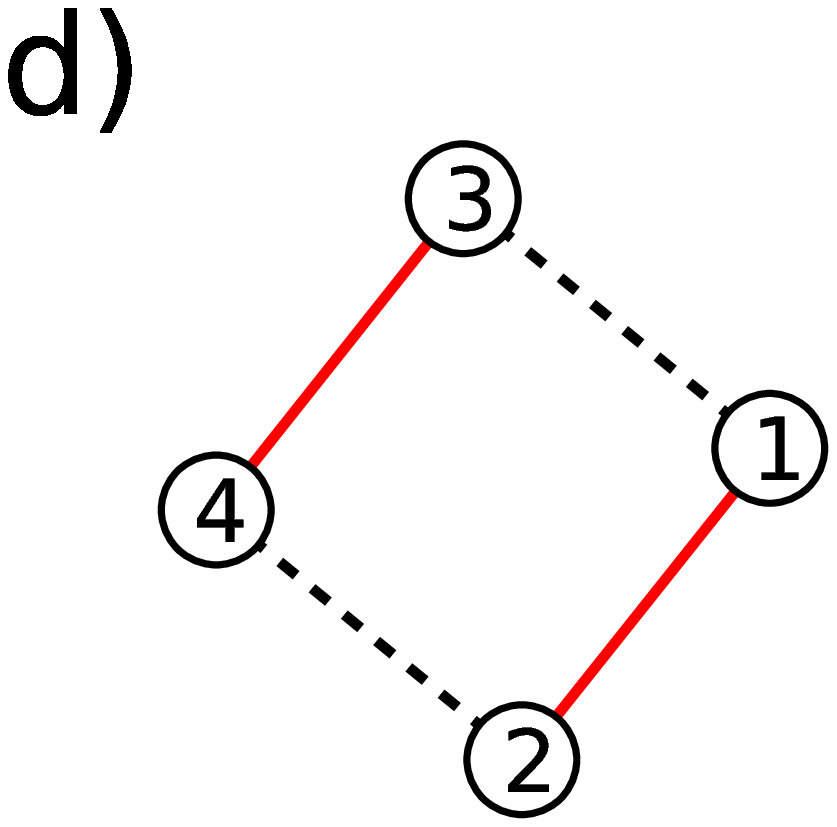}
\includegraphics[width=0.16\textwidth]{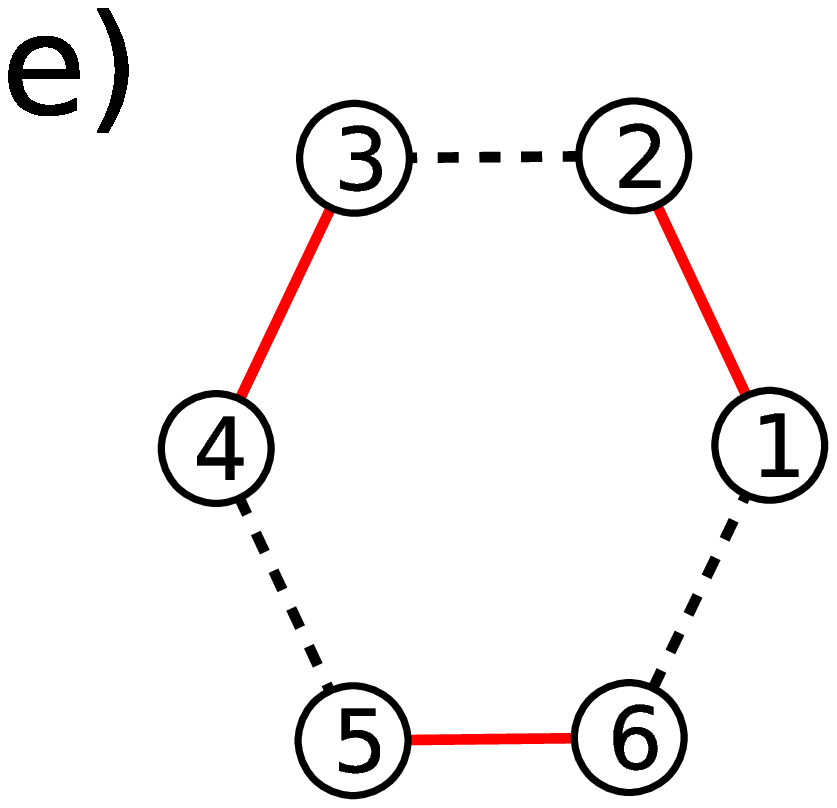}
\includegraphics[width=0.16\textwidth]{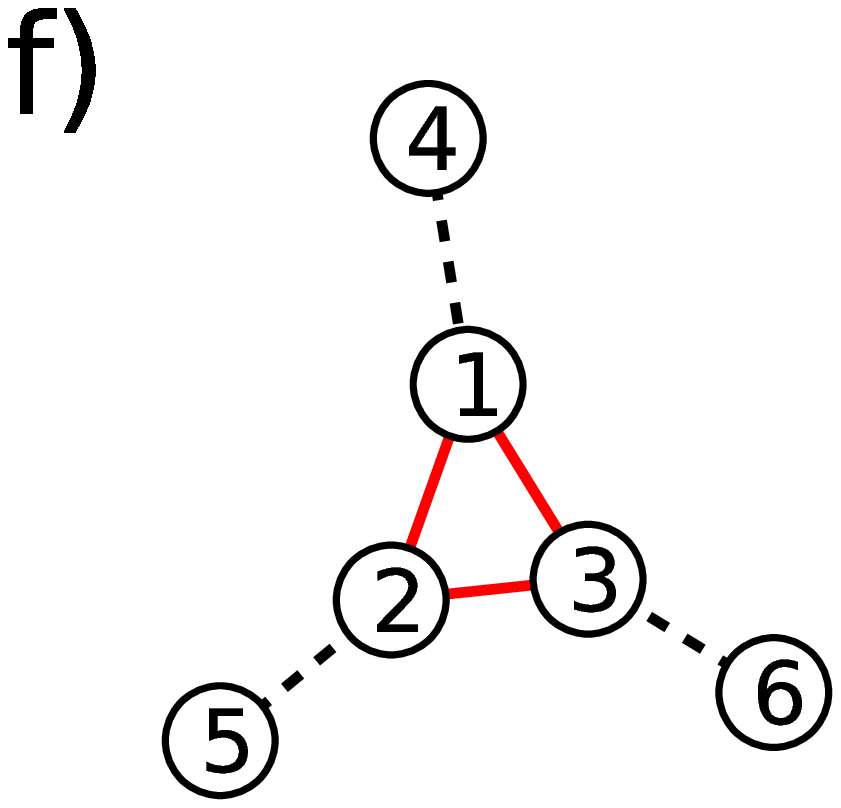}
\caption{Examples of two-colour interaction graphs, representing 
due to Eq. (\ref{U22}) different evolution operators:
 (a) $U = V_1 \in \mathrm{U}(n)$,
 (b) $U = W_{12} (V_1 \otimes V_2) \in \mathrm{U}(n^2)$, (c) $U = (W_{12} \otimes W_3)(V_1 \otimes V_{23})
 \in \mathrm{U}(n^3)$, (d) $U = (W_{12} \otimes W_{34})(V_{13} \otimes V_{24}) \in \mathrm{U}(n^4)$,
 (e) $U = (W_{12} \otimes W_{34} \otimes W_{56})(V_{23} \otimes V_{45} \otimes V_{61}) \in \mathrm{U}(n^6)$, 
(f) $U = (W_{123} \otimes W_4 \otimes W_5 \otimes W_6)(V_{14} \otimes V_{25} \otimes V_{36}) \in \mathrm{U}(n^6)$. 
Graphs (a) and (b) represent structureless (CUE) matrices, 
while graphs (c)--(f) correspond to structured matrices.}
\label{fig:graphs}
\end{figure}

Some examples of the two--colour graphs satisfying the cliques assumption
and representing evolution operators are shown in Fig.~\ref{fig:graphs}. 
The interactions are represented by either black dashed or red solid edges.
Observe that the system represented by the graph Fig.~\ref{fig:graphs}d
was already described by the former construction and shown in Fig.1.
On the other hand, the former approach, in which each
bond represents two subsystem, is not applicable for the system visualised in Fig.~\ref{fig:graphs}c which consists of an odd number of subsystems.

It is straightforward to generalize the above construction for three (or more) 
steps of the time evolution, which is determined by a graph
consisting of three (or more) classes of bonds.

\section{Spectral properties of graph-structured matrices}

Eigenvalues of a unitary matrix of order $N$ 
lie on the unit circle in the complex plane, 
so they have the form $z_j=e^{i\theta_j}$ for $j=1,\dots, N$.
 One can thus consider the probability density of the eigenphases, 
 $P(\theta)$, which is known to be uniform in $[0,2\pi)$
for random matrices of circular ensembles \cite{Mehta}.

Diagonalisation of a unitary matrix $U$ of order $N$ yields $N$ eigenphases.
After they are sorted, $\theta_1 \leq \theta_2 \leq \dots \leq \theta_N$, 
one may consider the normalized nearest neighbour spacing, 
\begin{equation}
S_i = \frac{N}{2 \pi} (\theta_{i+1} -\theta_i) .
\end{equation}
 Each Circular Ensemble of random matrices
(Poisson, orthogonal, unitary and symplectic)
 is characterized by the specific level spacing distribution $P(S)$. 
In the case of the unitary ensemble (CUE),
 equivalent to the Haar measure on the unitary group, 
one can use the Wigner surmise 
\begin{equation}
P(S) = \frac{32}{\pi^2} S^2 \exp\left(-\frac{4}{\pi} S^2\right),
\label{eq:PS_CUE}
\end{equation}
which is exact for random unitary matrices of order $2$ 
and gives a good approximation \cite{Mehta} also for large matrices, $N\to \infty$.

\begin{figure}
\centering
\includegraphics[width=0.32\textwidth]{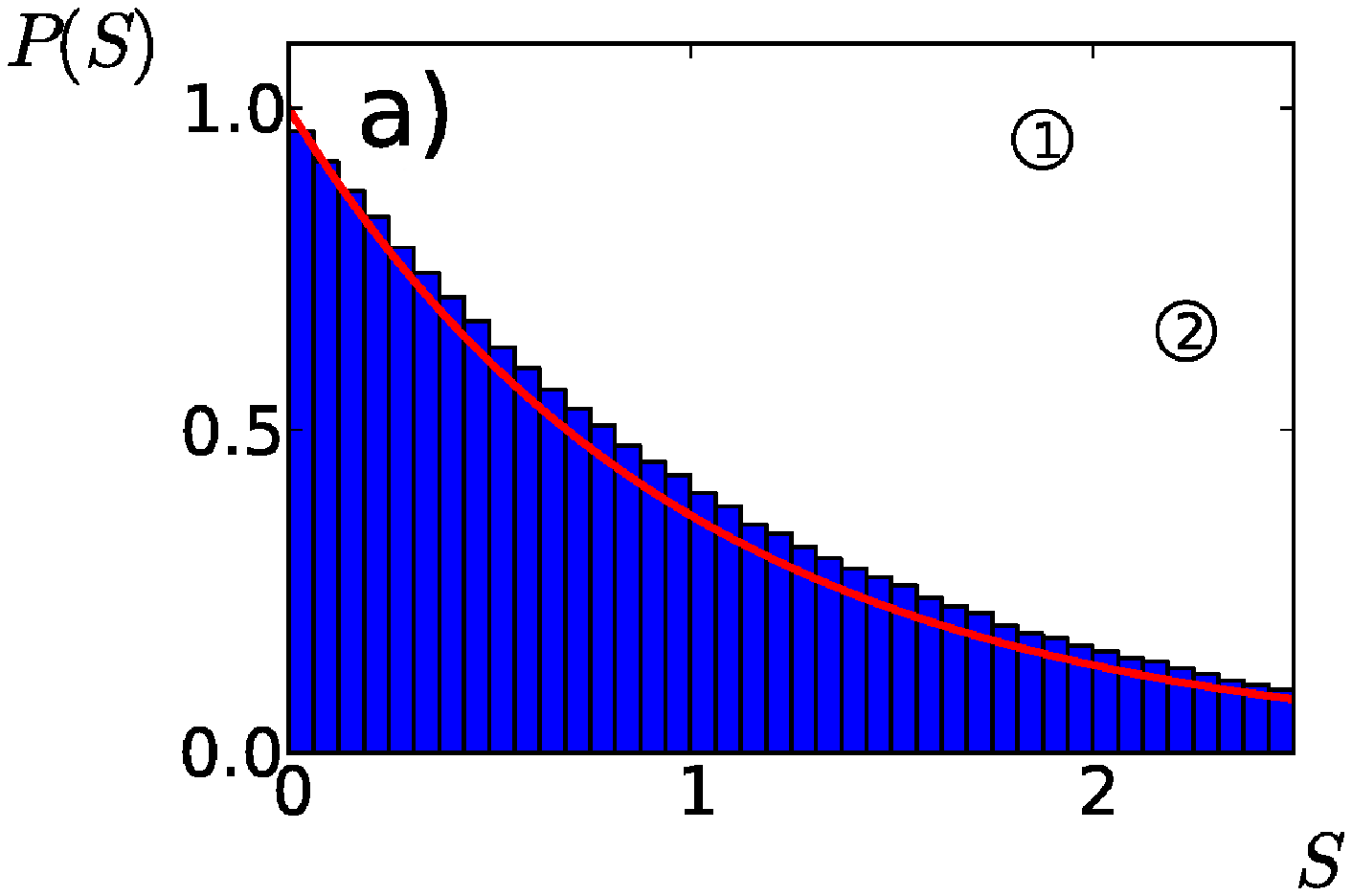}
\includegraphics[width=0.32\textwidth]{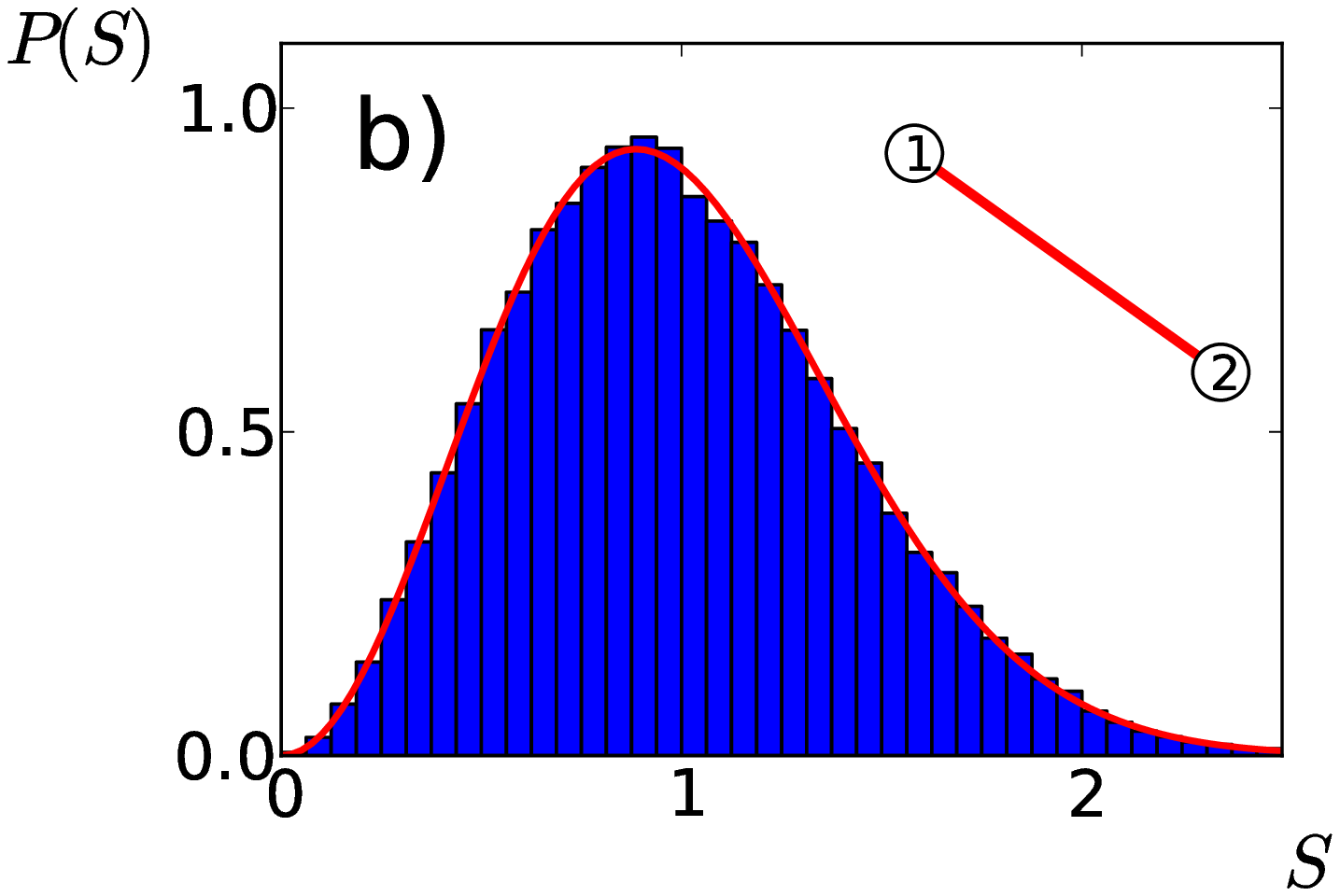}
\includegraphics[width=0.32\textwidth]{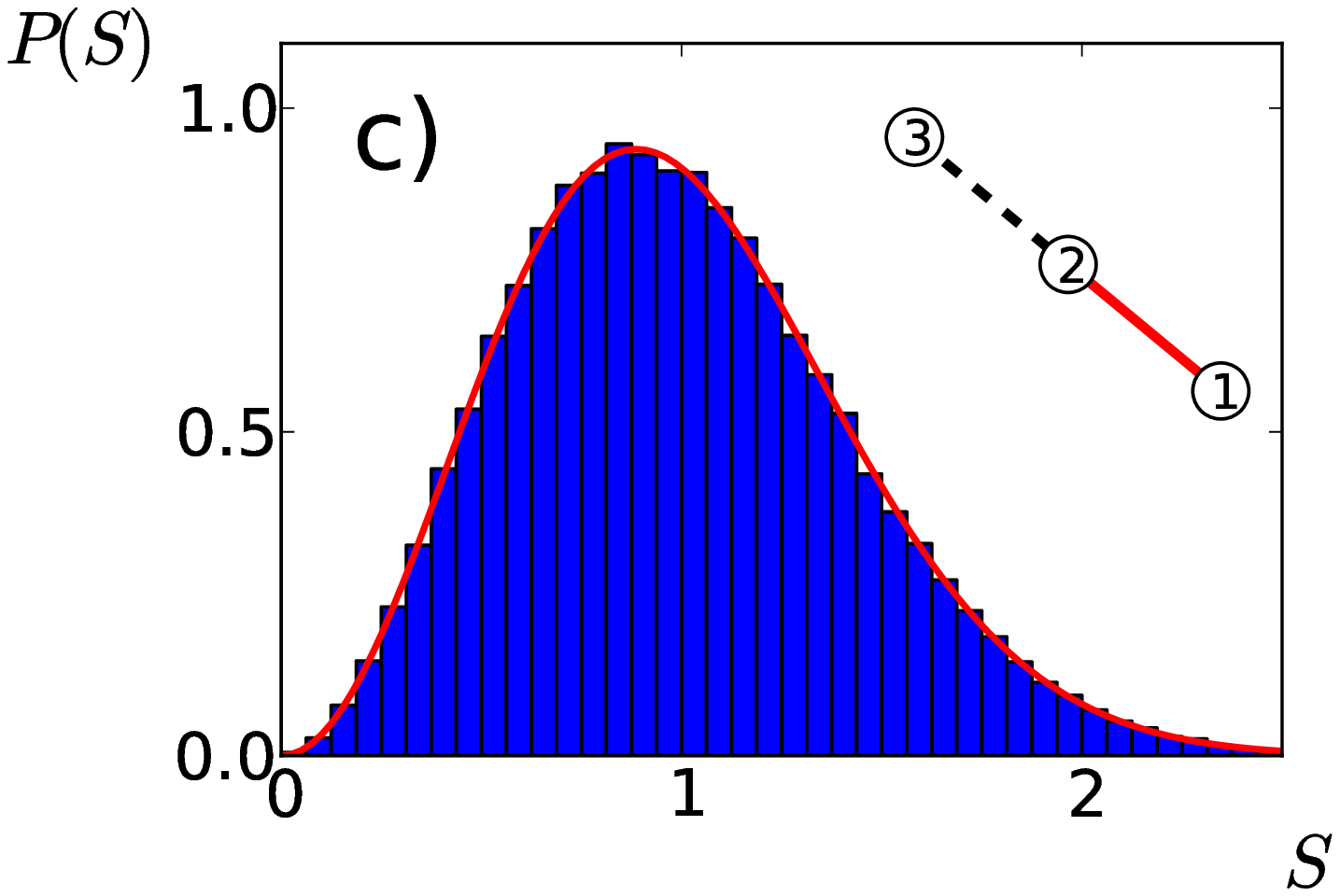}
\includegraphics[width=0.32\textwidth]{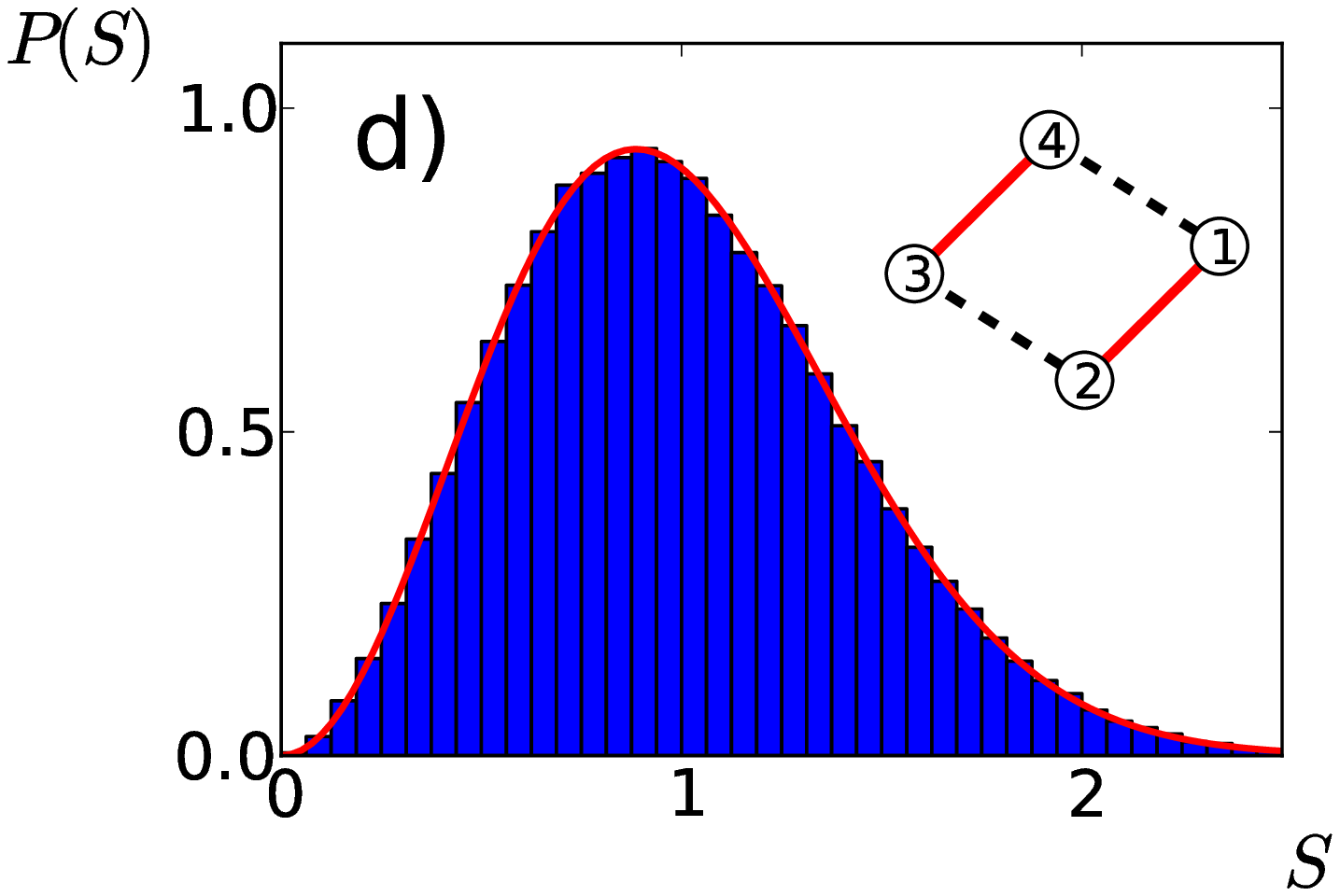}
\includegraphics[width=0.32\textwidth]{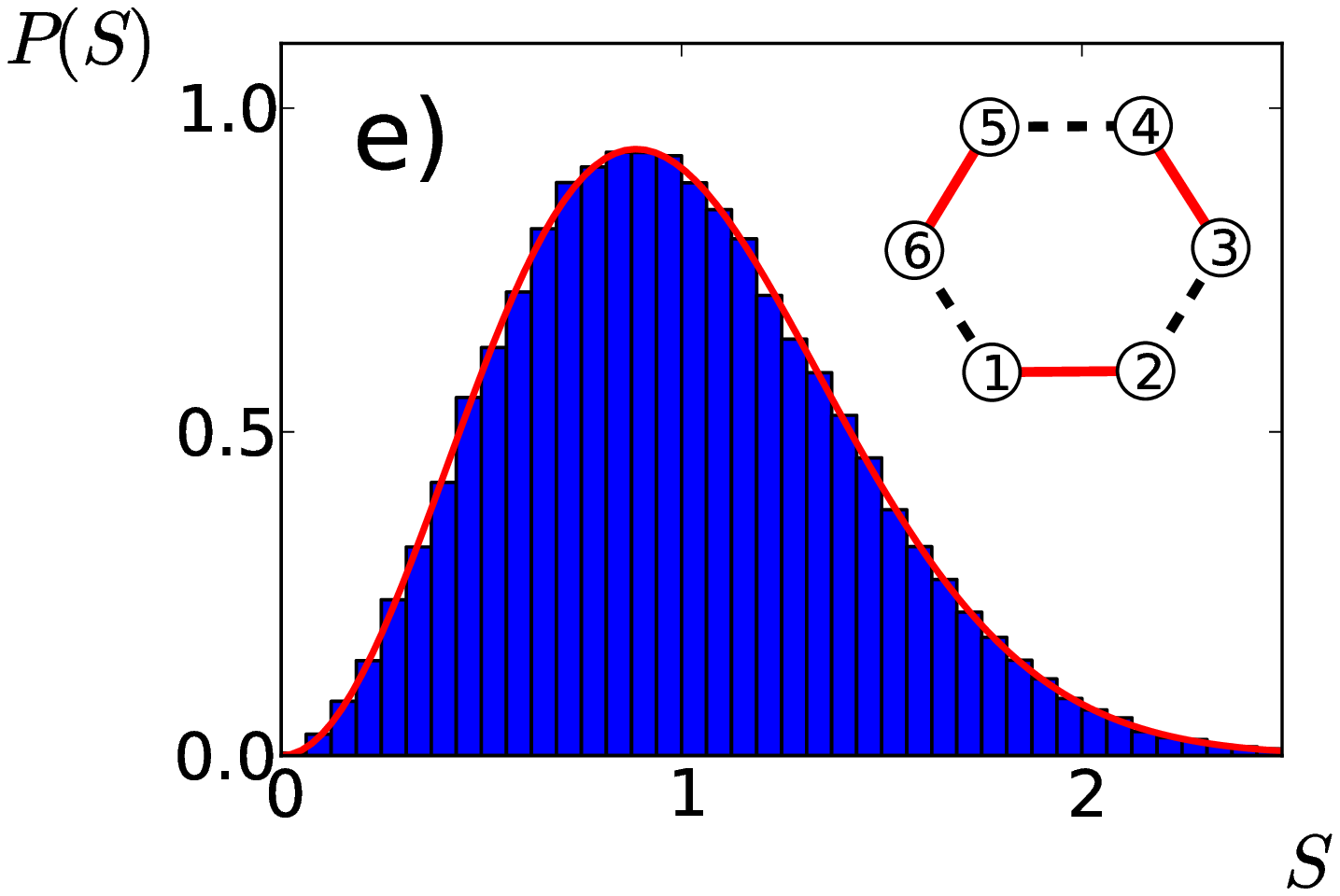}
\includegraphics[width=0.32\textwidth]{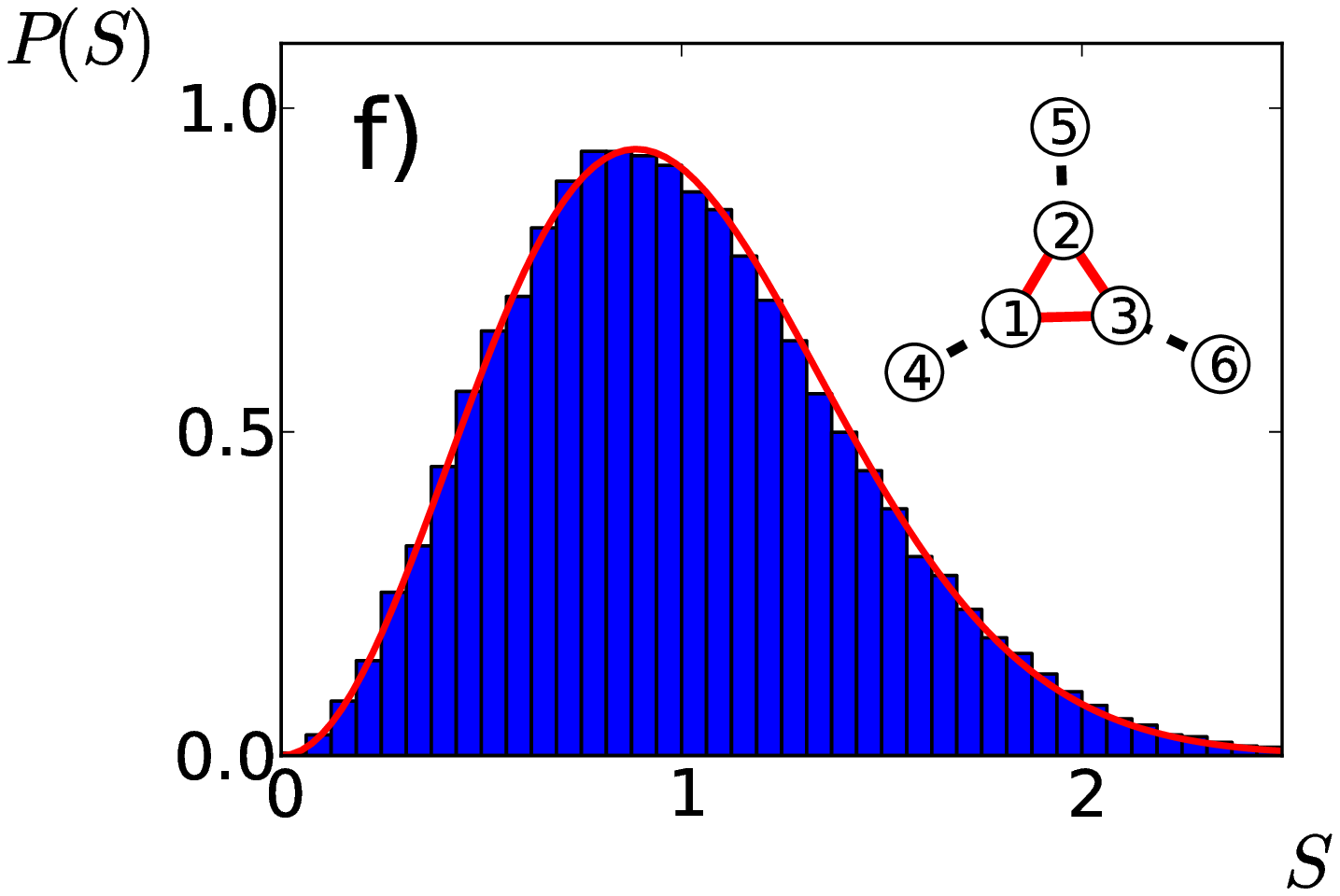}
\caption{Nearest level spacing statistics $P(S)$ for the unitary matrices 
corresponding to the graphs shown in the inset.
For a disconnected graph a) the matrices $U$ display Poissonian spectra, while
for connected graphs (b-f) the level spacing distribution is described by the Wigner 
distribution \eqref{eq:PS_CUE} characteristic of CUE. 
The matrix size $N$ reads 
(a) $N=256$, (b) $N=100$, (c) $N=1000$, (d) $N=2401$, (e) $N=729$, (f) $N=64$. 
In each case, the number of generated eigenvalues was of the order of $10^5$.}
\label{fig:PS}
\end{figure}

We were examining the spectral properties of various graph-structured matrices, built of smaller CUE matrices, 
which describe the interaction along the bonds or in the vertices.
For simplicity we will focus our attention here on systems which can be represented
by two-color graphs defined in Section \ref{sub2b}, and will use the graphical convention 
described in that section. We constructed numerically ensembles of
structured random unitary matrices corresponding to graphs presented in Fig. \ref{fig:PS}.
Random unitary matrices \cite{ZK94} used as building blocks of the construction
presented were obtained by the algorithm of Mezzadri \cite{Me07} - some details
concerning the numerical procedure are provided in the Appendix.
In all cases studied the level density $P(\theta)$ distribution is uniform,
which is the case for standard ensembles of random unitary matrices.

\medskip

 Regarding the structure of the graph\footnote{When we refer to (dis)connectivity of a graph, we disregard the fact that we actually have two (or more) different sets of bonds $B_i$. So in fact we are interested in the problem of connectivity of the graph $\Gamma=(Q, \cup_i B_i)$, where all the sets of bonds corresponding to the different stages of interactions have been summed.}, we distinguish two cases:
\begin{itemize}
\item {\bf The graph is disconnected.} In this case the matrix $U$ can be written as Kronecker (tensor) product of two or more smaller matrices. 
The resulting level spacing distribution is similar to Poissonian ensemble, for which 
the eigenvalues are uncorrelated. 

\item {\bf The graph is connected.} Our numerical results show, 
that several properties of the structured unitary matrices corresponding to connected graphs are similar to those of random structureless matrices.
In particular, if all the component matrices $W_\pi$ and $V_\pi$ are taken according to CUE, 
 the structured evolution matrix $U$ defined in Eq. (\ref{U22})
displays spectral properties characteristic to the Haar measure on $U(N)$ with $N=n^k$.
 Fig. \ref{fig:PS} presents spacing distributions $P(S)$ obtained
 for the matrices determined by exemplary graphs shown in each figure.
\end{itemize}

It is possible to relate these observations with recent results 
on tensor products of random unitary matrices, which display 
Poissonian level spacing in the limit of large matrices \cite{TSKZZ12,Tk13,STKZ13}.
Asymptotically both tensor product factors, 
$W=\otimes W_{\pi_1}$ and $V=\otimes V_{\pi_2}$ in Eq. (\ref{U22})
display thus a Poissonian spectra,
so the evolution operator represented in the eigenbasis of the first term 
has the form $Y^{\dagger}UY=P_1 XP_2X^{\dagger}$.
Here $P_1$ and $P_2$ denote diagonal unitary matrices
with Poissonian spectra of $W$ and $V$ respectively.
The unitary matrices $Y$ and $X$, are determined by eigenvectors of $W$ and $V$.

Note that unitary rotation matrix $Y$ does not influence the spectrum of $U$.
In the case of a connected graph the tensor products
defining $W$ and $V$ are taken with respect to different partitions,
hence it is possible to assume that the rotation matrix $X$
constructed out of eigenvectors has a CUE like properties \cite{Ha06}.
Hence we arrive at a composed ensemble of matrices \cite{PZK98},
of the form $U'=P_1XP_2X^{\dagger}$, which contains a product
of two diagonal matrices represented in different (random) bases.
Although both matrices $P_1$ and $P_2$ posses Poissonian level spacing distributions,
the composed ensemble display CUE-like spectra, which explains
the results obtained for matrices structured by connected graphs.
We performed numerical investigation for unitary matrices of size $N=100$
of this composed ensemble and found that level spacing distribution of 
$U'$ fits well to predictions of random matrices.

In the case of disconnected graph the both terms $W$ and $V$ in Eq. (\ref{U22})
have a tensor product structure with respect to the same partition,
for instance $U=(W\otimes W')(V \otimes V')=WV\otimes W'V'$.
Thus the evolution operator $U$ has a tensor product form, so
its level spacing distribution becomes asymptotically 
Poissonian \cite{TSKZZ12}, which explains properties of 
matrices structured by disconnected graphs.

The above arguments work asymptotically for evolution operators
describing the interaction represented by two-colour connected graphs,
such that each vertex contains at least one connection of each colour.
To analyze to what extend this assumption can be relaxed
we investigated an $L$ step interaction, described by a larger class of graphs
with bonds of $L$ different colours. In the simplest case, consider a chain
of $L+1$ subsystems, such that in each moment only two neighbouring subsystems
are involved: in the first step the first subsystem interacts with the second,
in the second step the interaction couples subsystems two and three, an so on.

\begin{figure}
\centering
\includegraphics[width=0.32\textwidth]{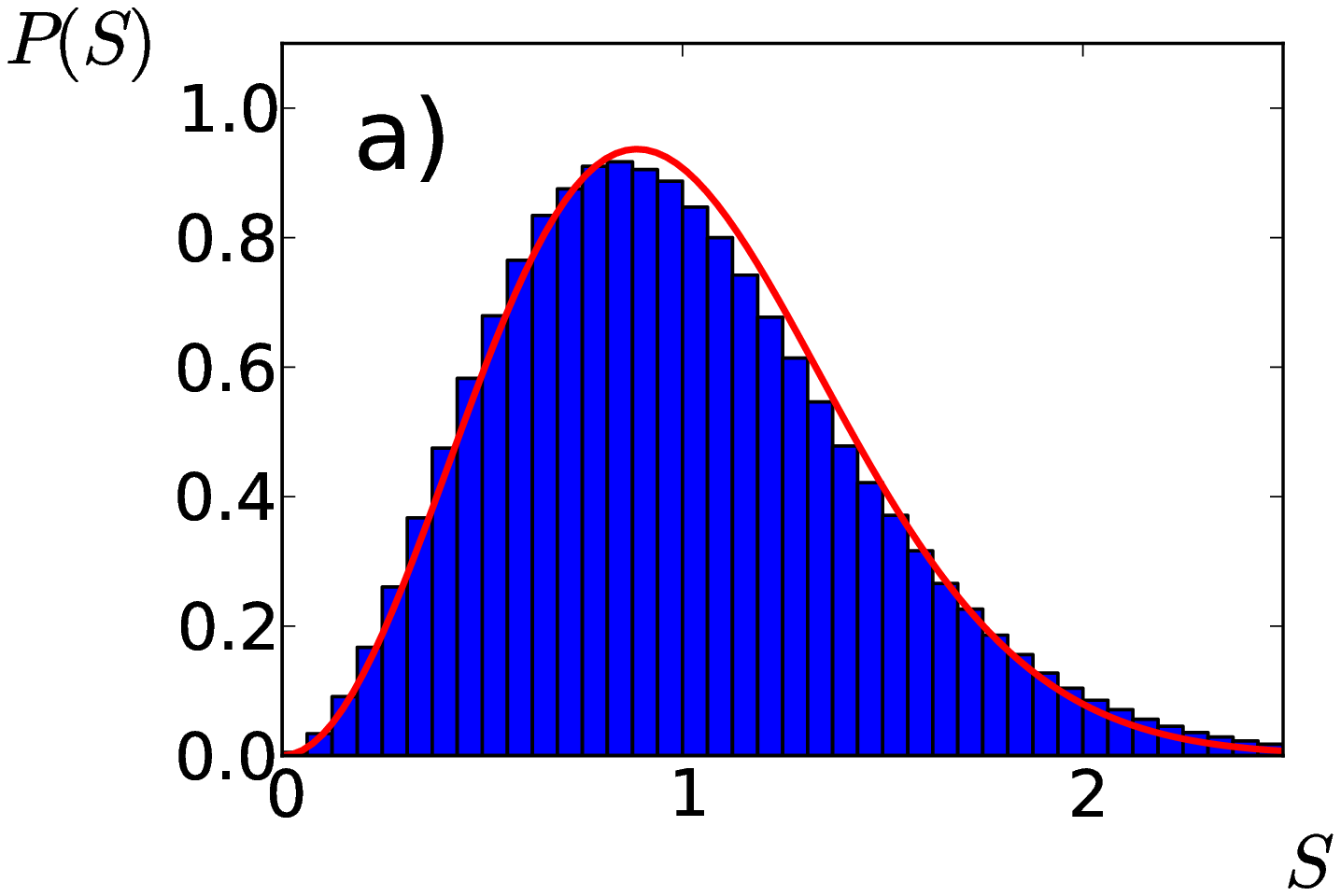}
\includegraphics[width=0.32\textwidth]{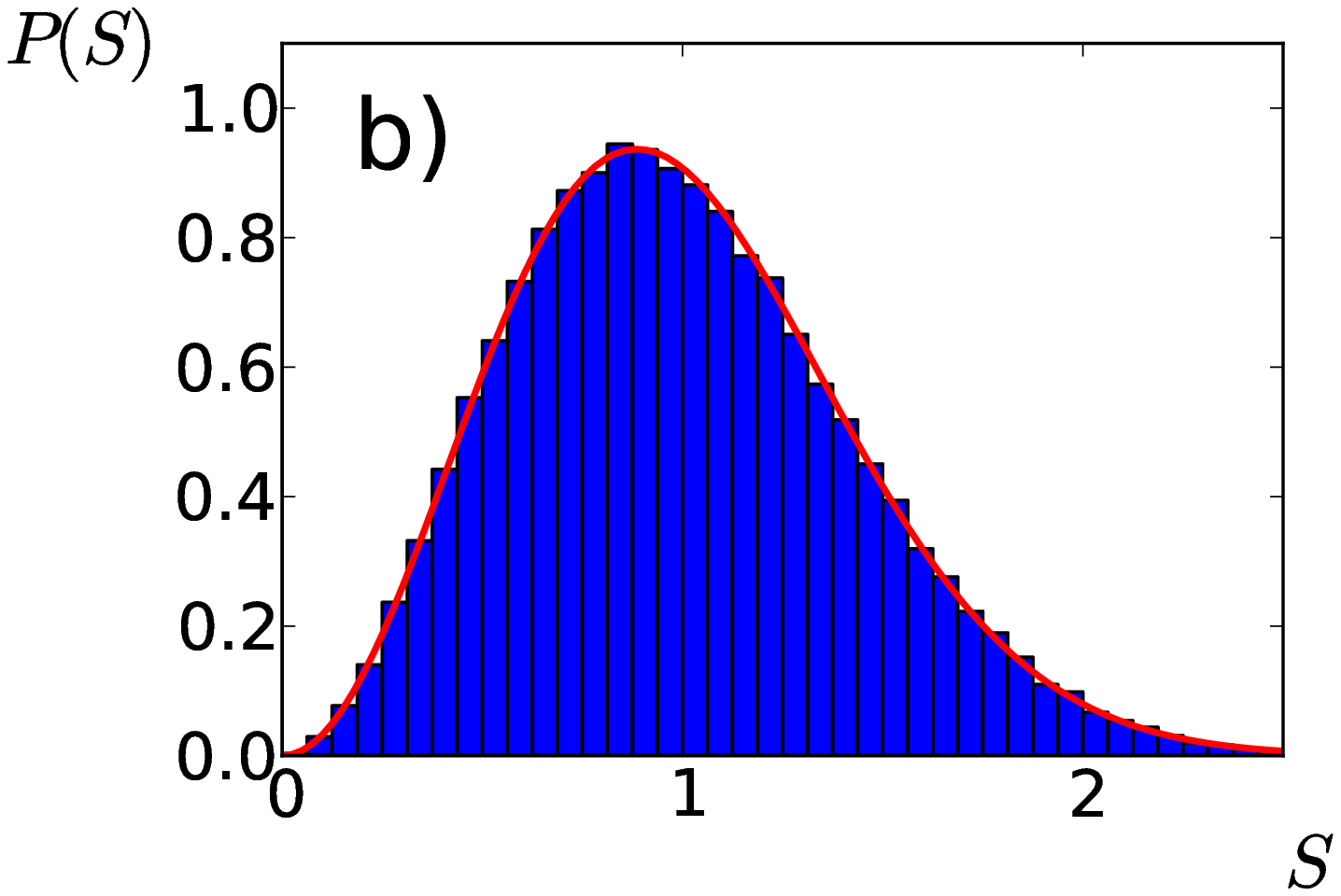}
\caption{Level spacing distribution $P(S)$ for unitary matrices $U$ corresponding to a five-step evolution of a six--chain consisting of (a) six qubits, $n=2$ and (b) six qutrits, $n=3$. 
The sample size is (a) $20000$ and (b) $400$ matrices.
In the $i$-th step of the evolution only $i$-th and $(i+1)$-th subsystem
interact so $U$ is given by Eq. (\ref{eq:5steps}).
}
\label{fig:5step_PS}
\end{figure}

This interaction can be represented by a chain of $L+1$
vertices, such that all $L$ bounds between the neighbouring vertices are
of a different colour, associated with the interaction in specific time steps.
Figure \ref{fig:5step_PS} shows the level spacing distribution 
for such a model with $L=5$ time steps,
\begin{equation}
U = WVXYZ = (W_{12} \otimes W_3 \otimes W_4 \otimes W_5 \otimes W_6) \cdots (Z_1 \otimes Z_2 \otimes Z_3 \otimes Z_4 \otimes Z_{56}) .
\label{eq:5steps}
\end{equation}
 In the case of six qubit system
($n=2$ -- see Fig. \ref{fig:5step_PS}a) some deviations 
from the CUE results are visible,
while in the case of six qutrits ($n=3$, Fig. \ref{fig:5step_PS}b)
the distribution follows predictions of random matrices with a good accuracy.
Hence Wigner--like spectral properties of the evolution operator
can be obtained under very week assumptions on the interaction,
as the first and the last subsystems are coupled only indirectly
by the $L$--step interaction.

\section{Eigenvectors of graph--structured unitary matrices}

Matrices of eigenvectors of random unitary matrices are known to be 
distributed according to the Haar measure on the unitary group \cite{Mehta}.
It is interesting to analyze statistical properties of eigenvectors 
of unitary matrices associated with a given graph 
and compare them with predictions for random matrices.

Let us write the eigenequation of a unitary matrix as
$U\ket{\chi_j}= e^{i \vartheta_j}{\ket{\chi_j}}$. 
The eigenstates are normalized
$ \braket{\chi_j}{\chi_j} = 1$ for $j=1,\dots, N$,
so the complex expansion coefficients $\chi_{ji}$ of the state
$|\chi_j\rangle$ in the computational basis
satisfy $\sum_{i=1}^N |\chi_{ji}|^2=1$ for $j=1,\dots, N$.
These $N$ non-negative numbers form thus a probability vector,
the distribution of which can be characterized by its Shannon entropy.

\subsection{Eigenvector entropy}

For any unitary matrix $U$ one defines the eigenvector entropy as
the average Shannon entropy of a single eigenvector
\begin{equation}
H_{ev}(U) \equiv -\dfrac{1}{N} \sum_{i,j} |\chi_{ji}|^2 \log |\chi_{ji}|^2.
\label{eq:Eev_def}
\end{equation}
Let us note, that this quantity coincides with the 
entropy of the unistochastic matrix \cite{ZKSS03} 
corresponding to the unitary matrix of eigenvectors of $U$.
The mean entropy of eigenvector or a random unitary matrix of order $N$,
coincides with the mean entropy of a random complex vector \cite{Jo90,BZ06}
\begin{equation}
  \langle H_{el}\rangle  = \psi(N+1)-\psi(2) = \sum_{j=2}^N \frac{1}{j} 
  \label{mean}
\end{equation}
Here $\psi(x)$ denotes the digamma function, $d \ln \Gamma(x) / dx$.

\begin{figure}
\centering
\includegraphics[width=0.42\textwidth]{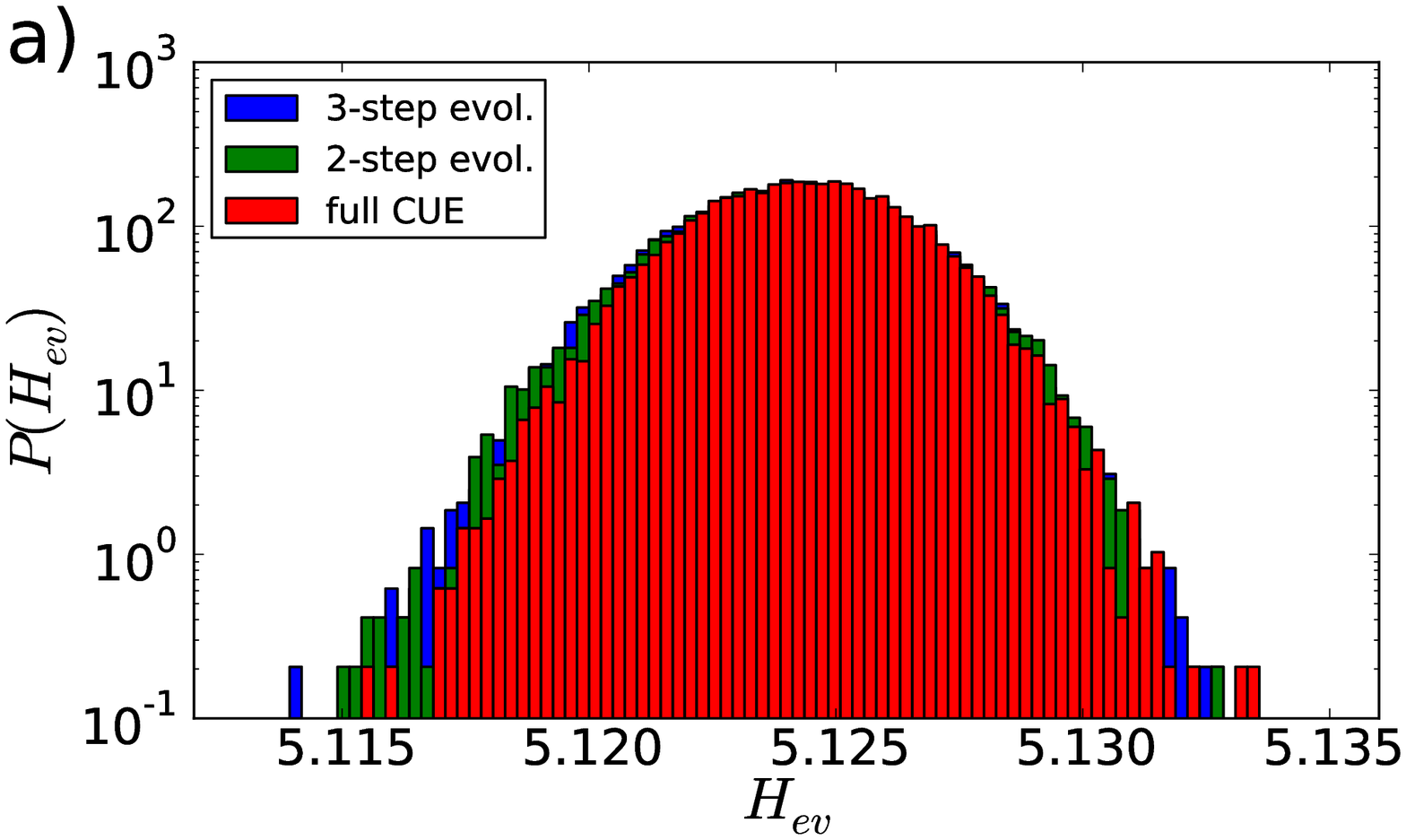}
\includegraphics[width=0.42\textwidth]{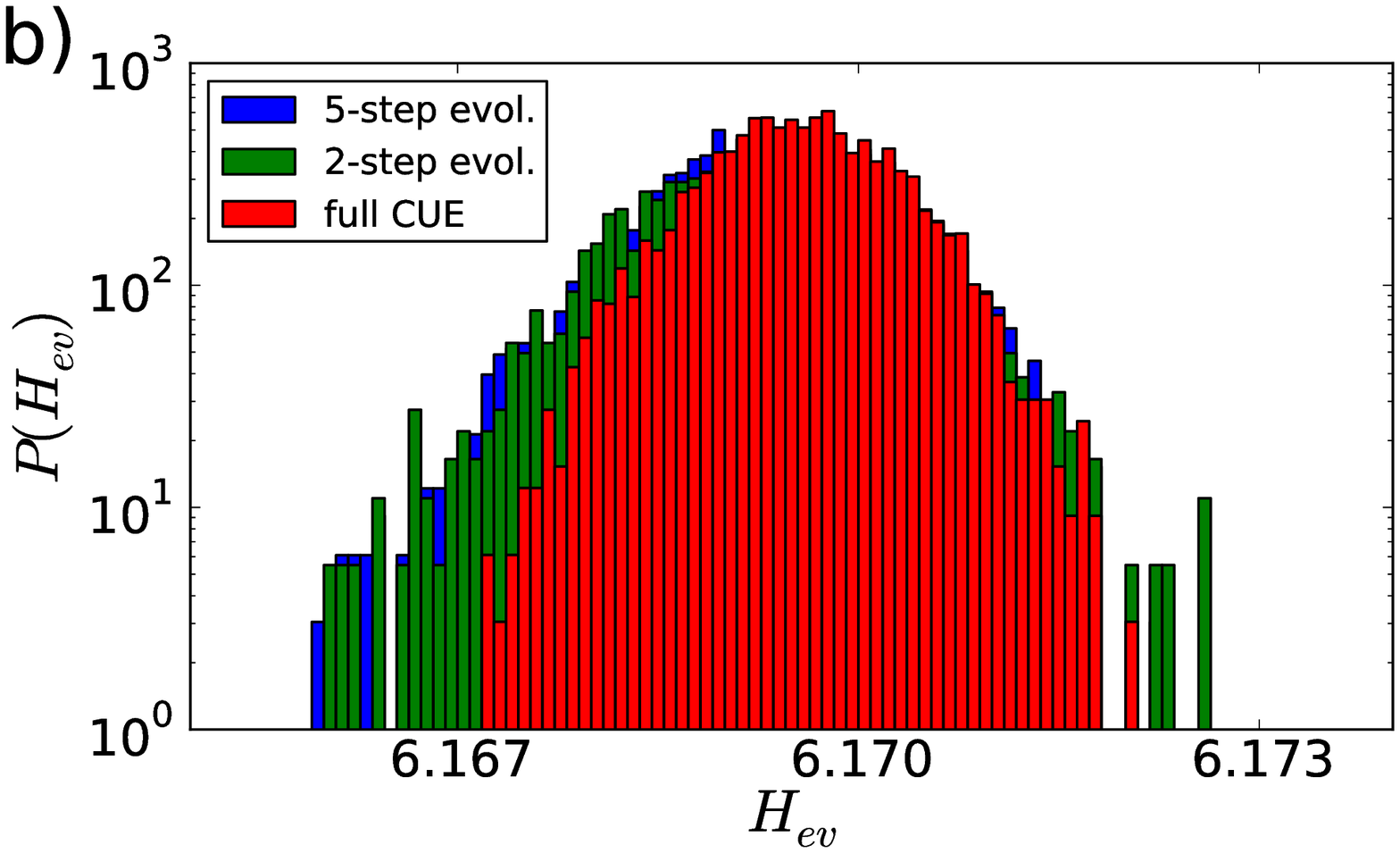}
\caption{Distribution of eigenvector entropy $P(H_{ev}(U))$ 
for unitary matrices $U$ associated to a) 
a chain consisting of $4$ subsystems, 
 $U = WV = (W_{12} \otimes W_{34}) (V_1 \otimes V_{23} \otimes V_4)$
realized for $n=4$, sample size $20000$ (green) and the corresponding three--step evolution (blue)
and b) six--particle chain with a two--step local interaction
 given by Eq. (\ref{eq:5steps})
obtained for $n=3$, sample size $50000$ (green), 
and its five--step generalization (blue). The larger dimension $n$ 
of a single system,
the better agreement with the CUE data (red) obtained for the same total dimension $N$.}
\label{fig:5step_evec-ent}
\end{figure}

A comparison of the distribution of eigenvector entropy
for graph--structured and unstructured random unitary matrices 
is presented in Fig. \ref{fig:5step_evec-ent}.
Our numerical investigations show that statistical properties
of eigenvectors of unitary matrices associated to 
connected graphs coincide for large dimensions 
 with the prediction of random CUE matrices.

\subsection{Entropy and purity of a reduced state}

A unitary matrix $U$ associated to a graph acts on Hilbert space with a tensor product
structure and corresponds to composed systems. Let us divide the system into two parts,
labeled by $A$ and $B$.
The eigenvectors $\ket{\chi}$ of $U$ can be considered 
as pure states of a bi--partite system $AB$.
One can thus investigate their {\sl entanglement entropy}
with respect to the partition $A-B$, equal to 
the von Neumann entropy $H(\sigma)=-{\rm Tr} \sigma \ln \sigma$
of the reduced state, $\sigma_A={\rm Tr}_B|\chi\rangle \langle \chi|$.

For random vectors of the size $N_A N_B$
the average entropy of a subsystem of dimension $N_A$ reads \cite{Pa93,ZS01}
\begin{equation}
\langle H \rangle \approx \log N_A - \dfrac{N_A-1}{2N_B}
\label{eq:page_conj}
\end{equation}
Here it is assumed that the dimension $N_A$ of the reduced state 
is large and $N_B \ge N_A$ so that the maximal entropy $H_{max}$ is equal to 
$\log N_A$. If both subsystems are equal $N_B=N_A=\sqrt{N}$, the 
reduced states $\sigma_A$ are distributed uniformly according to the
Hilbert--Schmidt measure in the set of mixed quantum states
and the average entropy is 
$\langle H \rangle_{\rm HS} \approx \dfrac{1}{2} \log N - \dfrac{1}{2}$.
By definition it is equal to the average entanglement entropy
of random pure states of size $N=N_A^2$.

\medskip 

In order to provide an alternative characterization 
of degree of mixing of a quantum state one often uses {\sl purity},
Tr$\sigma^2 $, equal to unity for a pure state.
This quantity applied to a reduced state 
$\sigma_A={\rm Tr}_B|\chi\rangle \langle \chi|$,
carries information about entanglement of a bi--partite state $|\chi\rangle$.
For pure random states of size $N=N_A N_B$ the average purity is 
\cite{Lu78,BZ06} 
$\langle R \rangle = \dfrac{N_A + N_B}{N_A N_B + 1}$.

\begin{figure}
\centering
\includegraphics[width=0.4\textwidth]{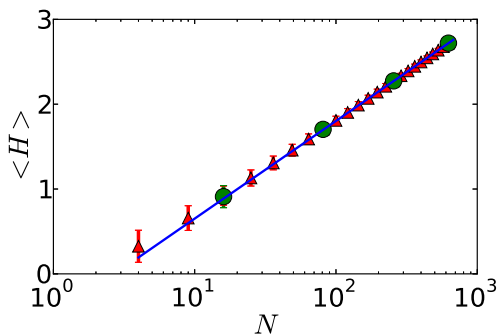}
\caption{Mean entanglement entropy $\langle H \rangle$ of the eigenvectors of a structured unitary matrix $U$, versus its size $N$. The red triangles denote the system of two particles, Fig. \ref{fig:graphs}b, and the green circles --- four particles, Fig. \ref{fig:graphs}d. In all the cases the dimensions of the two subsystems $A$ and $B$ are equal,  $N_A=N_B = \sqrt{N}$.
The solid line follows from the prediction for CUE matrices (Eq. (\ref{eq:page_conj}), with $N_A=N_B = \sqrt{N}$).}
\label{fig:entropy}
\end{figure}

\begin{figure}
\centering
\includegraphics[width=0.4\textwidth]{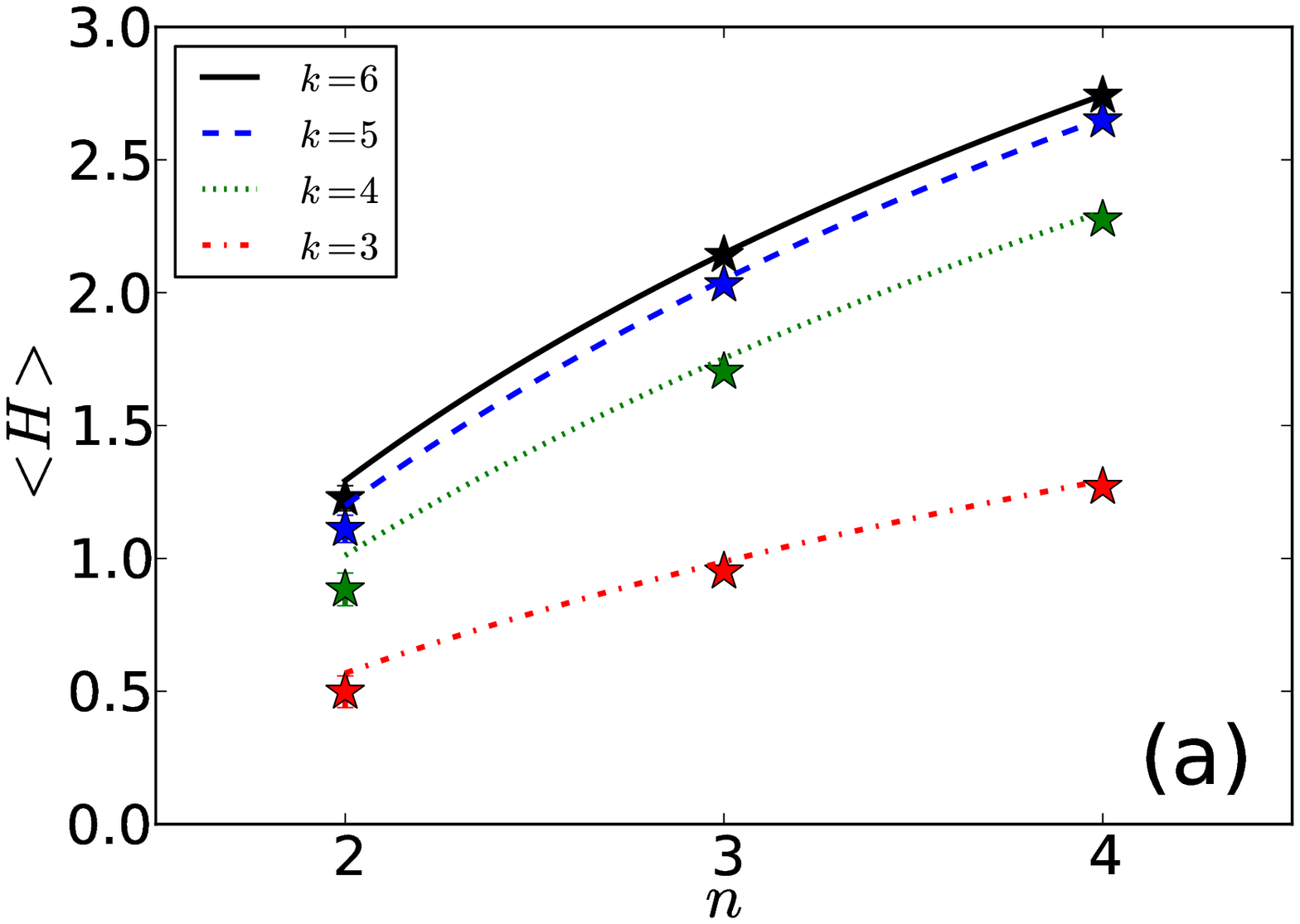}
\includegraphics[width=0.4\textwidth]{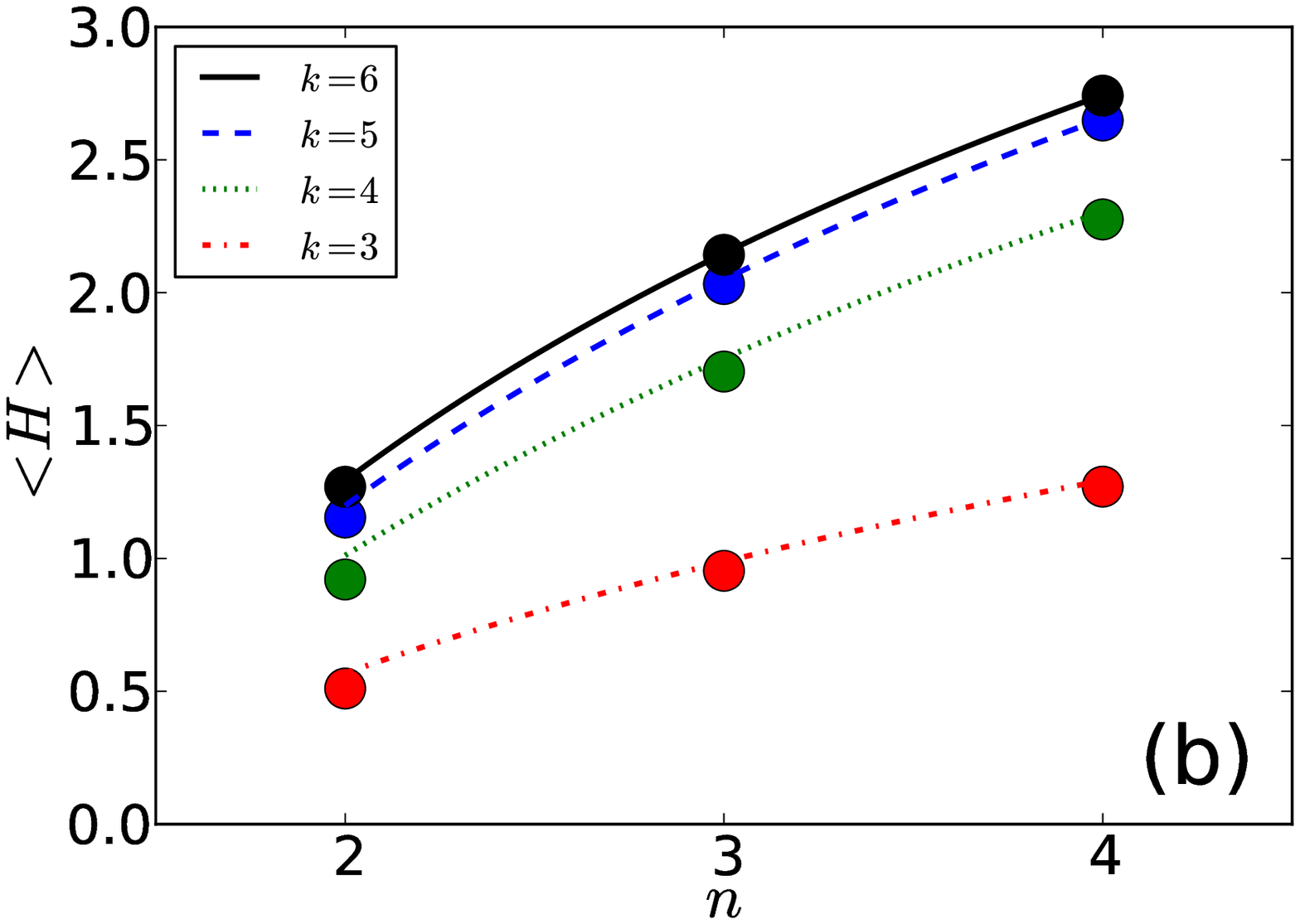}
\caption{Linear chains of subsystems of length $k=3,4,5,6$.
(a) Average entropy $\langle H \rangle$ of the subsystem consisting of 
two peripheral particles, versus the dimension $n$ of the 
Hilbert space of a single particle (symbols) 
and predictions of random matrices following from (\ref{eq:page_conj}) -- lines.
 (b) Comparison with the non-structured random unitary matrices 
of order $N=n^k$.}
\label{fig:chains_entropy_of_ends}
\end{figure}

Average entanglement entropy of eigenvectors of random unitary matrices
associated to exemplary graphs are shown in 
Fig. \ref{fig:entropy} and Fig. \ref{fig:chains_entropy_of_ends}
while the average purity is presented in Fig. \ref{fig:entropy_purity_cxc}.b.
In the case of connected graphs the entropy of entanglement and
purity of the reduced eigenvectors of associated unitary matrices
coincide thus with properties of random vectors described by random matrices.

\subsection{Projection to a smaller subspace}
As eigenvectors $|\chi\rangle$ of $U$ denote pure states of 
a multipartite systems, it is of interest to investigate properties of their
projections onto a smaller subspace. 
We have performed such a procedure for a system of three particles (see Fig. \ref{fig:graphs}c), with unequal dimensions of the subspaces, 
$N_A=N_C \ne N_B$.
 After choosing one of the basis vectors $\ket{i_B}$ of $\mathcal{H}_B$, 
the eigenstates $|\Psi_{ABC}\rangle=|\chi\rangle$
were projected onto the subspace $\mathcal{H}_A \otimes \mathcal{H}_C$,
\begin{equation}
\ket{\tilde\Psi_{AC}} \equiv \braket{i_B}{\Psi_{ABC}}.
\label{proje}
\end{equation}
Average entanglement between particles $A$ and $C$ was characterized 
by the purity and the von-Neumann entropy of the reduced state
$\sigma_A = \Tr_C \ketbra{\tilde\Psi_{AC}}{\tilde\Psi_{AC}}$.
The averaging is performed over all $N_B$ basis vectors in the
 Hilbert space $\mathcal{H}_B$ as well as over several realizations
of the corresponding random matrix.

\begin{figure}
\centering
\includegraphics[width=0.3\textwidth]{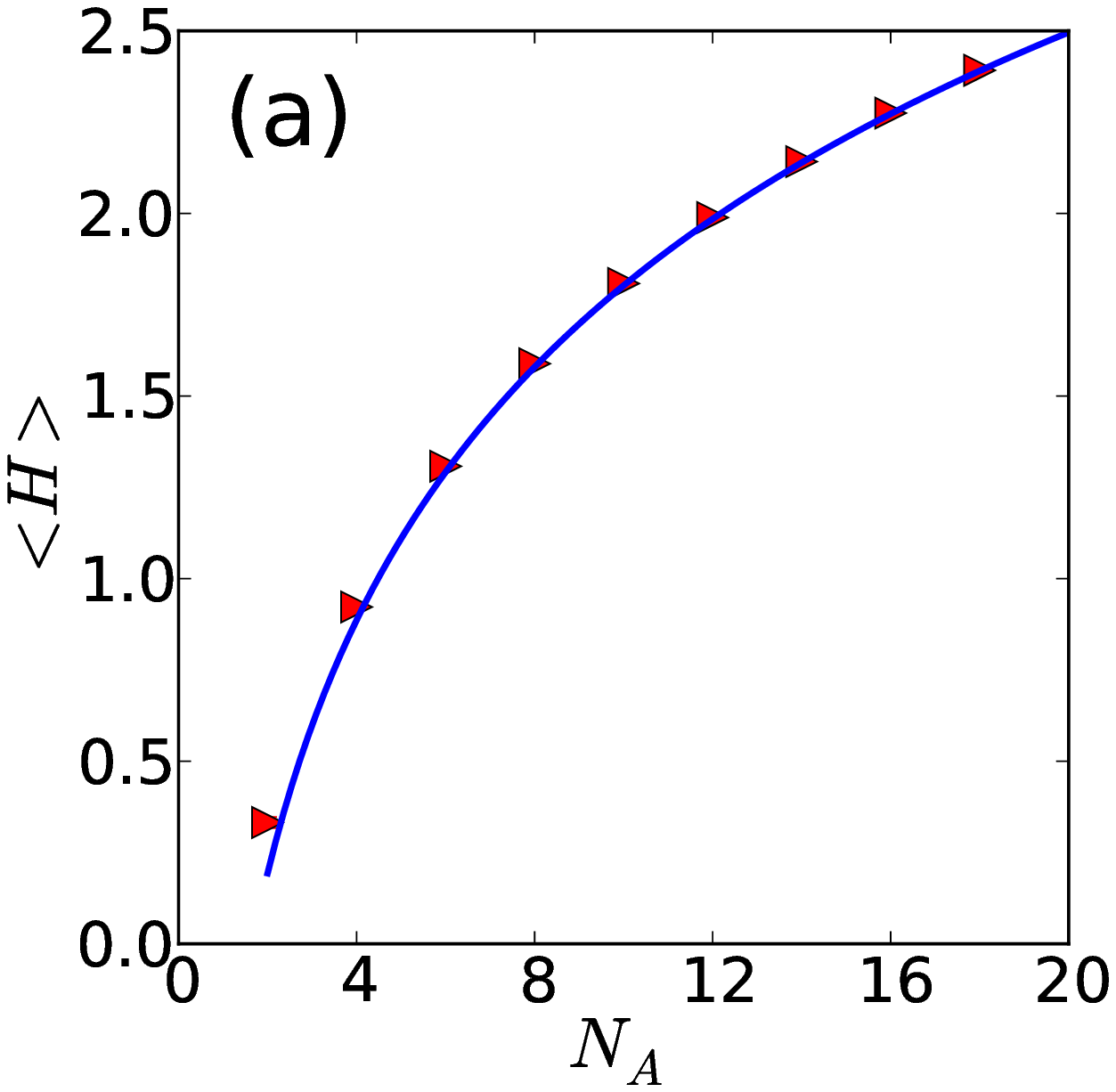}
\includegraphics[width=0.3\textwidth]{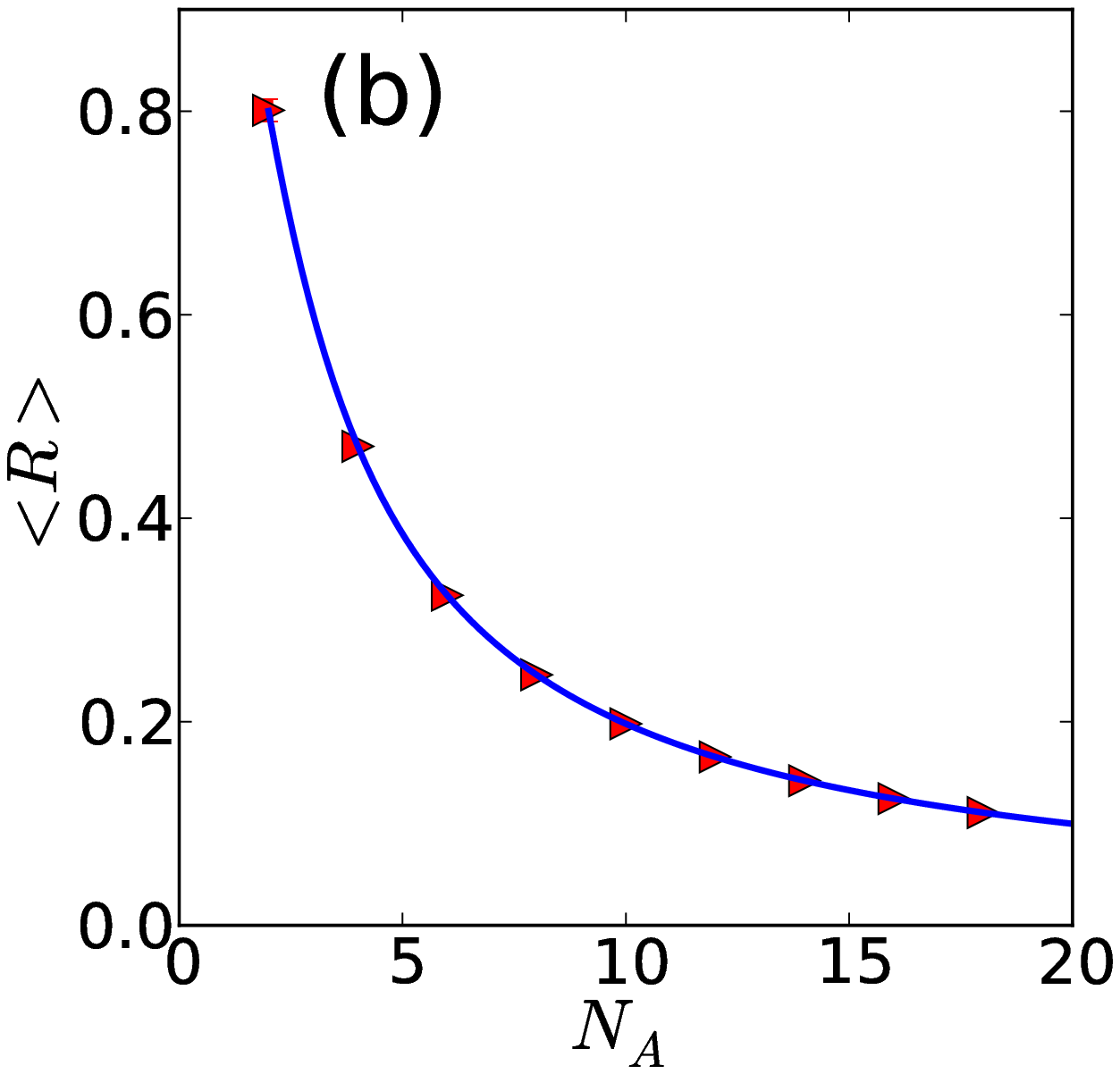}
\caption{Properties of eigenvectors of matrices associated to a three--chain projected onto a single subspace as in Eq. (\ref{proje}) 
 as a function of its dimension $n=N_A$:  a) mean entropy  $\langle H \rangle$ and b) mean purity  $\langle R \rangle $ compared 
with predictions of random matrices.}
\label{fig:entropy_purity_cxc}
\end{figure}

Interestingly, entanglement between the peripheral particles does not depend on the dimension of the central particle, which acts as a proxy of interactions. 
As shown in Fig. \ref{fig:entropy_purity_cxc}, entropy and purity
of the projected eigenvectors exhibit the CUE-like behavior.

\section{Distribution of matrix elements}

 Although unitary matrices related to connected graphs display 
 statistical properties of spectra and eigenvectors according
 to predictions of random matrices, the distribution 
     of their elements is different.
 To describe quantitatively the distribution of entries of $U$  we use 
  the {\it element entropy},
\begin{equation}
H_{el}(U) \equiv -\dfrac{1}{N} \sum_{ij} |u_{ij}|^2 \log |u_{ij}|^2.
\label{eq:Eel_def}
\end{equation}
For any unitary matrix with a tensor product structure
the element  entropy is additive, 
\begin{equation}
H_{el}(U\otimes V) = H_{el}(U) + H_{el}(V).
\end{equation}

\begin{figure}
\centering
\begin{tabular}{p{0.32\textwidth}p{0.32\textwidth}p{0.32\textwidth}}
a) & b) & c) \\ 
\end{tabular} 
\includegraphics[width=0.32\textwidth]{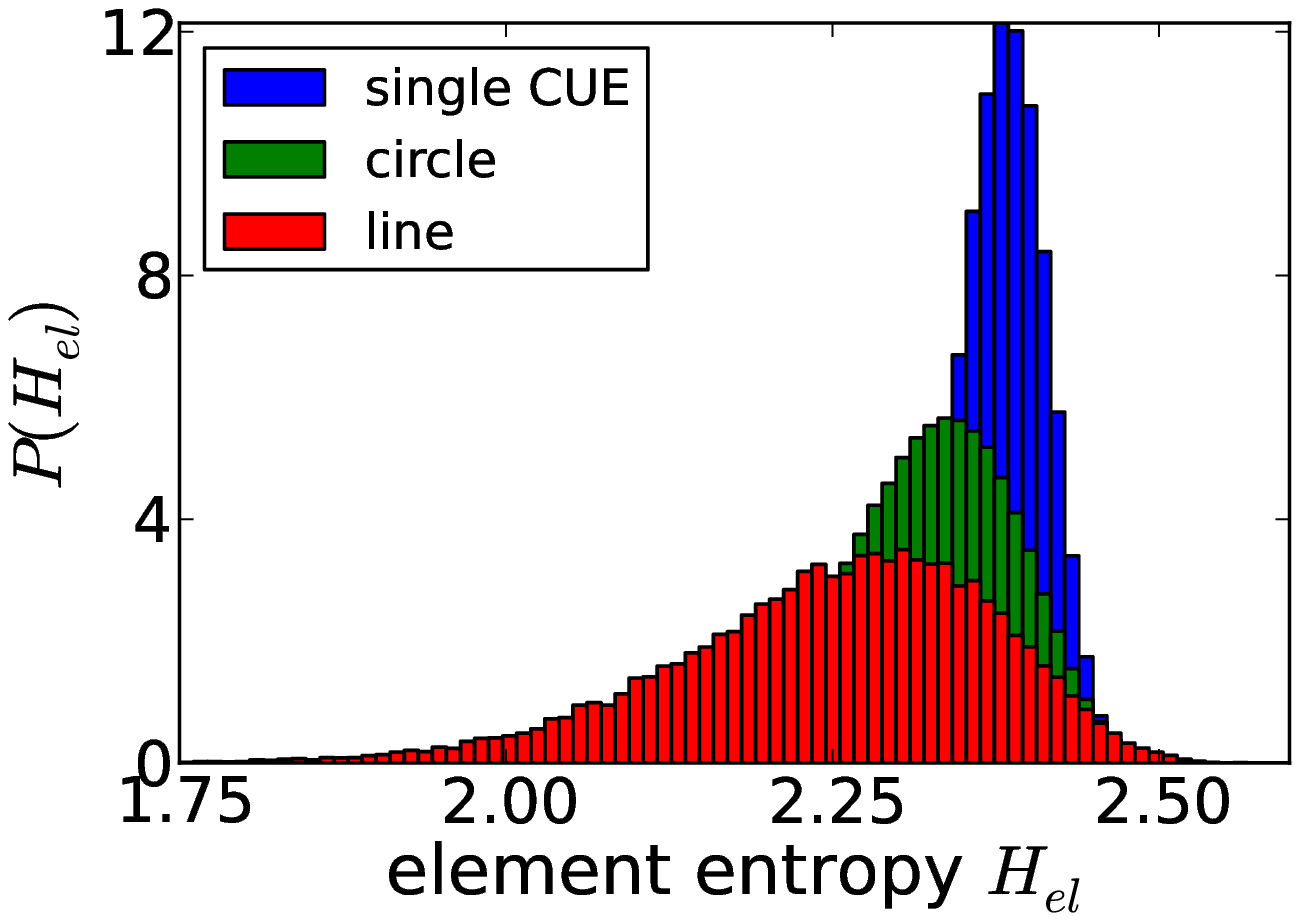}
\includegraphics[width=0.32\textwidth]{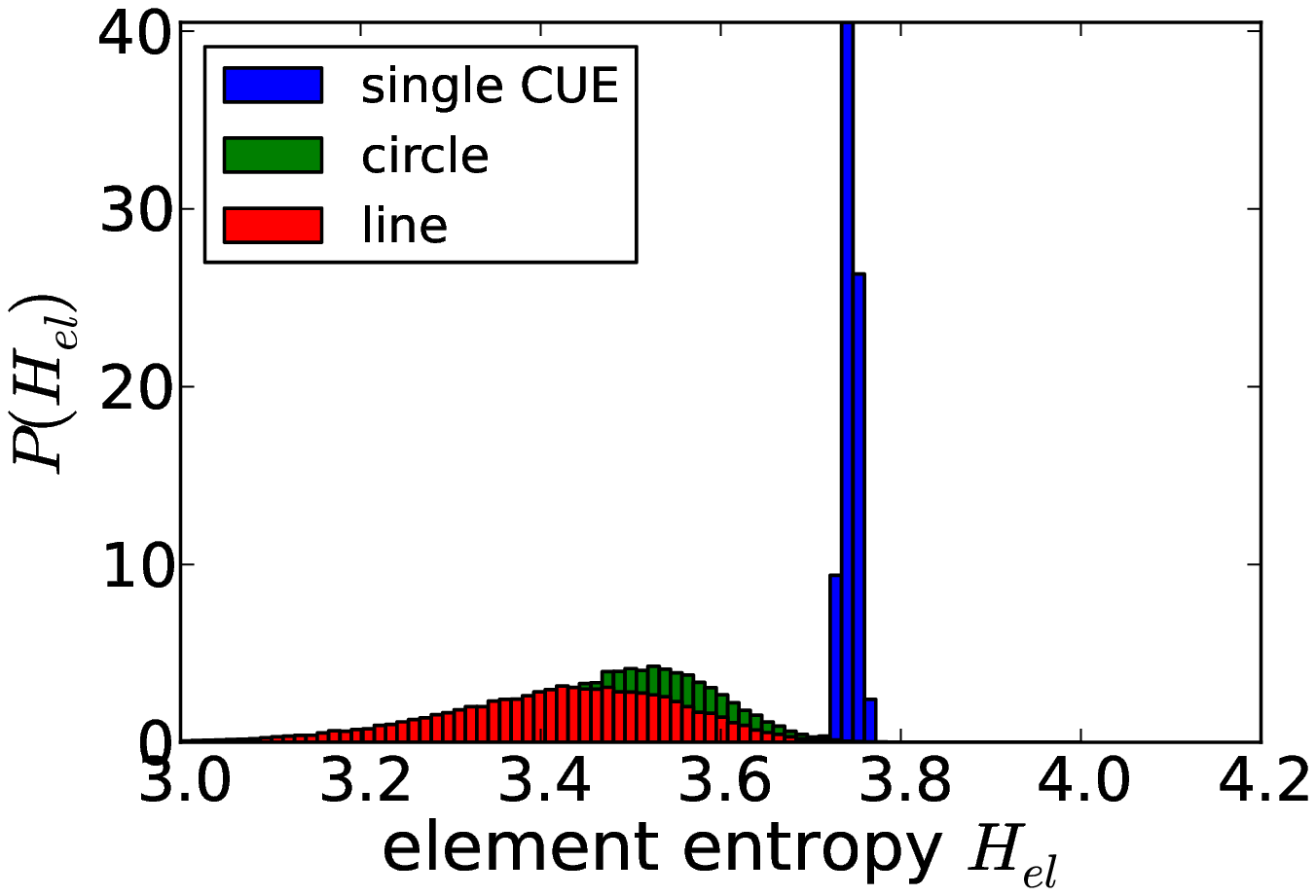}
\includegraphics[width=0.32\textwidth]{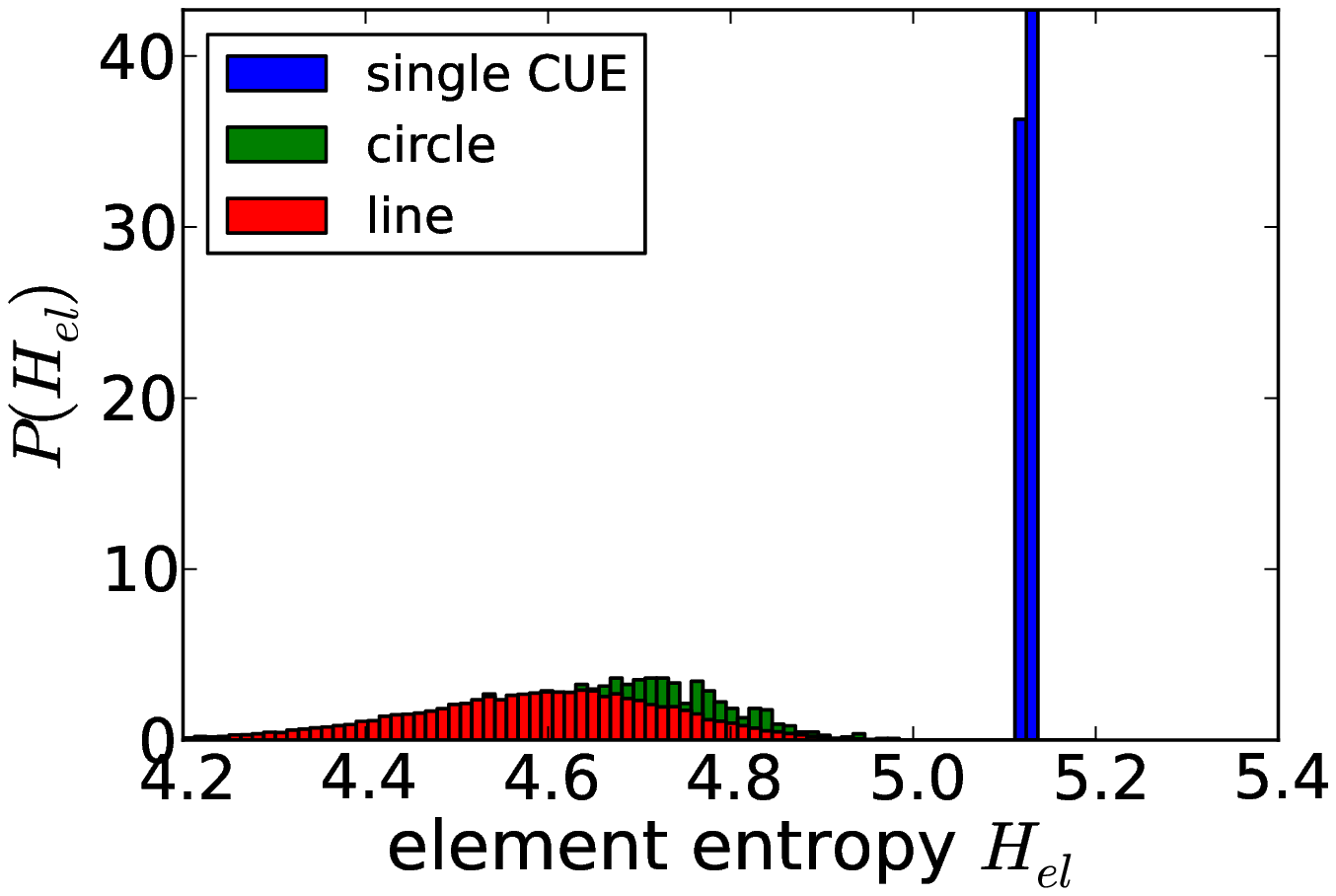}
\caption{Distribution of the {\it element entropy} $H_{el}$
for random matrices associated to a chain (such as in Fig. \ref{fig:graphs}c; red histogram), circle (such as in Figs. \ref{fig:graphs}d or \ref{fig:graphs}e; green)
and random CUE matrices (blue). Dimension of a
single subsystem is $n=2$, while the number of particles
is  (a) $k=4$ (b) $k=6$ (c)~$k=8$.}
\label{fig:el_entropy_dist}
\end{figure}

The distribution of the element entropy for
random matrices associated to exemplary connected graphs
is shown in  in Figs. \ref{fig:el_entropy_dist}.
Any construction of a CUE matrix of order $N$ requires more
independent random numbers than to generate smaller matrices
necessary to construct matrices associated to a graph.
Therefore the distribution of the element entropy is narrower
for CUE matrices than for matrices associated to a graph
and it allows one to distinguish between 
structured and unstructured random matrices.

\section{Concluding remarks}

We proposed a construction of random  unitary matrices which, 
interpreted as representations of the Hamiltonian evolution operators, 
correspond to various scenarios of the interaction between quantum objects. 
The structure of such matrices can be determined by a graph
in two different ways. In the first method each bond represents two 
interacting subsystems, while the other phase of the interaction
couples all subsystems connected by a single vertex.
In the alternative method the vertices represent individual subsystems 
and the edges represent  interaction between them.
In this work we focus on the case, in which no details concerning 
the interaction Hamiltonian are known, so the interaction is modeled by  
random unitary matrices generated according to the Haar measure
on the unitary group.

Our numerical results support the conjecture  that the spectral properties 
of structured unitary matrices associated to a connected graph  
are well described by predictions of CUE random matrices.
 This concerns the level density  $P(\theta)$,
 the nearest neighbour spacing distribution $P(S)$,
and also statistical properties of eigenvectors.
On the other hand, the simplest method do distinguish
between ensembles of structured unitary matrices and CUE
is to analyze statistics of their elements and compare e.g.
the element entropy.

Statistical properties of elements of random matrices associated to 
a graph $\Gamma$  are determined by its topology.
Analyzing statistics of the elements and the traces of unitary matrices 
associated to two different graphs in several cases 
one can distinguish between these two graphs. However,
in general it seems not to be possible
to determine in this way, whether two investigated graphs are isomorphic.

Making use of unitary random matrix associated to a  graph $\Gamma$
one can act with it on a given separable pure state 
and arrive at a random pure state, 
$|\Psi\rangle=U|1 \otimes \cdots \otimes 1\rangle$.
In this way one obtains an ensemble
of random states associated to a graph $\Gamma$, 
which can be considered as a generalization
of the ensembles investigated in \cite{CNZ10,CNZ13}.
On the other hand the above construction of an ensemble of quantum states 
corresponding to a graph differs from the notion of quantum graphs studied  
by Gnutzmann and  Smilansky \cite{GS06}
or these related to microwave experiments investigated in \cite{HBPZS04,LBHBS11}.

As the standard model discussed in this work corresponds to a 
two--step interaction, represented by a two--colour graph,
it can be generalized for an arbitrary number of $L$ time steps,
and the corresponding $L$--colour graphs.
It is worth to emphasize that even in the case
 of the chain graphs, with
local interaction coupling two neighbouring sites only,
CUE--like spectral properties are observed for large system size
provided the number  of the time steps $L$ is sufficiently large.
This means that the randomness can be transfered by a step-wise
nearest neighbour interaction, from the first subsystem to the last one.

Random unitary matrices play an important role
in various protocols of quantum information processing.
However, the number of quantum gates necessary to implement
a random unitary matrix  grows exponentially with the
number of qubits. \cite{EW+03}. As a substitute one 
may construct various pseudo random matrices, the statistical
properties of which should be similar to these of CUE \cite{EW+03,Zn07,We13}.

The scheme of random unitary matrices associated with a graph,
developed in this work, can be thus directly applied
to construct a large random unitary matrices 
out of a few much smaller  unitary matrices of order.
Consider for instance random matrices
\begin{equation}
U \ = \  V  \; W \ = \ 
\bigl(   V_{2,3}\otimes V_{4,5} \otimes \dots \otimes 
V_{k,1} \bigr) 
\bigl(  W_{1,2} \otimes W_{3,4} 
\otimes \dots \otimes W_{k-1,k} \bigr)
\label{chain}
\end{equation}
associated with a two colour ring of $k-1$ subsystems
interacting with one neighbour in phase one (bonds of a first colour)
and with the other on phase two (bonds of the second colour).
This natural extension of a six-partite system
shown in Fig. \ref{fig:graphs}e
allows one to obtain random unitary matrices of size $N=n^k$
out of $k$ smaller CUE matrices of size $n^2$.
As discussed in the appendix the generation time of such graph-structured matrices 
can be shorter compared to the standard CUE sampling.

\medskip

It is a pleasure to thank Benoit Collins, Marek Ku{\'s}, Ion Nechita and Tomasz Tkocz for fruitful discussions and helpful remarks.
Financial support by the Polish National Science Centre under the contract number DEC-2011/01/M/ST2/00379 
and by the Deutsche ForschungsGemeinschaft under the project
SFB Transregio--12 is gratefully acknowledged.

\section*{Appendix: Time costs of the matrix generation}

The calculations were performed on two PCs, with processors Intel(R) Core(TM)2 CPU T5500 @ 1.66 GHz and Intel(R) Xeon(R) 
CPU X3430 @ 2.40 GHz. The computer program was written in C++ and used the Armadillo \cite{Sa10} linear algebra library. 
The sampling of CUE was implemented according to the algorithm proposed by Mezzadri \cite{Me07}, 
based on the QR decomposition. In a single run, $T$ matrices were generated and then diagonalized to obtain the eigenvalues.

\begin{figure}[htbp]
\centering
\includegraphics[width=0.49\textwidth]{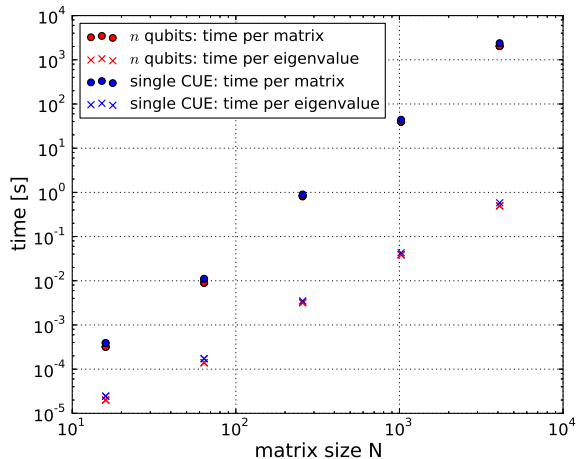}
\caption{Time costs of the matrix generation: line-structured system of $n$ qubits, see Fig. \ref{fig:graphs}b, (red points) 
vs direct construction of random CUE matrices (blue points).}
\label{fig:n_qubits_gen_time}
\end{figure}

As shown in Fig. \ref{fig:n_qubits_gen_time}
and Table \ref{tab:gen}, construction of random matrices associated to a graph
costs less time than the corresponding CUE matrix obtained by the 
Mezzadri algorithm.
However, to investigate spectral properties of a matrix obtained
one needs to diagonalise it, and for a larger matrix size $N$ the 
computing time needed to construct a random matrix
is dominated by the diagonalization time.

\begin{table}[htbp]
\centering
\begin{tabular}{|c|c|c|c|}
\hline matrix type & \# matrices & time [s] & rel. time [\%]  \\ 
\hline
\hline CUE, $N=256$ & 1000 & 85.93 & 100 \\ 
\hline square graph, $4$ matrices of size $n=4$
& 1000 & 47.24 & 55.0 \\ 
\hline
\hline CUE, $N=625$ & 100 & 100.27 & 100 \\ 
\hline square graph, $4$ matrices of size $n=5$
& 100 & 62.26 & 62.1 \\ 
\hline 
\end{tabular}
\caption{Comparison of the time required to generate random matrices
associated to a square graph and obtained by a CUE algorithm.
 Processor: Intel(R) Xeon(R) CPU X3430 @ 2.40 GHz.}
\label{tab:gen}
\end{table}


\end{document}